Review Article

# Radiomics and artificial Intelligence for thyroid cancer diagnosis: Concepts, challenges, and solutions

**Running title: Radiomics and Thyroid cancer**

**Milad Yousefi [1, #], Hadi Vahedi [2, #, *], Shadi Farabi Maleki [3], Mahya Ahmadpour Youshanlui [4], Aida Jafari [5], Parisa Rostami [6, 7], Kais I. Abdul-Lateef Al-Abdullah [8], Ryszard Tadeusiewicz [9], Paweł Pławiak [10, 11], Roohallah Alizadehsani [12], Siamak Pedrammehr [13]**

[1]*Faculty of Mathematics, Statistics and Computer Sciences, University of Tabriz, Tabriz, Iran.*

[2]*Research Center for Evidence-Based Health Management, Maragheh University of Medical Sciences, Maragheh, Iran*

[3]*Nikookari Eye Center, Tabriz University of Medical Sciences, Tabriz, Iran.*

[4]*Immunology Research Center, Tabriz University of Medical Sciences, Tabriz, Iran.*

[5]*Department of Traditional Medicine, TITU University, Turkey Branch of Belgium, Van, Turkey.*

[6]*Research Center for Evidence-Based Medicine, Iranian EBM Centre: A JBI Centre of Excellence, Faculty of Medicine, Tabriz University of Medical Sciences, Tabriz, Iran.*

[7]*Student research committee, Tabriz university of medical sciences, Tabriz, Iran.*

[8]*Department of Electrical Engineering, Australian University (AU), Mubarak Al-Kabeer 1411, Kuwait.*

[9]*AGH University of Science and Technology, Department of Biocybernetics and Biomedical Engineering, Krakow, Poland.*

[10]*Department of Computer Science, Faculty of Computer Science and Telecommunications, Cracow University of Technology, Warszawska 24, 31-155 Krakow, Poland.*

[11]*Institute of Theoretical and Applied Informatics, Polish Academy of Sciences, Bałtycka 5, 44-100 Gliwice, Poland*

[12]*Faculty of Medicine, Queensland University of Technology, Brisbane QLD Australia.*

[13]*Faculty of Design, Tabriz Islamic Art University, Tabriz, Iran.*

[#]*These authors contributed equally to this work*

# These authors contributed equally to this work and should be regarded as co–first authors.

***Corresponding Author**:

Hadi Vahedi, MD

*Research Center for Evidence-Based Health Management, Maragheh University of Medical Sciences, Maragheh, Iran*

*Email:* Hadivahedimrg@gmail.com *vahedih@tbzmed.ac.ir*, https://orcid.org/0009-0008-4769-451X

Postal address: East Azerbaijan Province, Maragheh, North Moallem Street, Maragheh University of Medical Sciences

Postal code: 5516773111
Tel: +989901210821

| Name | Degree | Email | ORCiD |
|---|---|---|---|
| Milad Yousefi | MSc | miladdyousefi@gmail.com | https://orcid.org/0009-0006-5790-5144 |
| Hadi Vahedi | MD | hadivahedimrg@gmail.com | https://orcid.org/0009-0008-4769-451X |
| Shadi Farabi Maleki | MD | shadfa8@gmail.com | https://orcid.org/0009-0003-6124-2599 |
| Mahya Ahmadpour Youshanlui | MD | mahyaahmadpouryoushanlui@gmail.com | https://orcid.org/0000-0003-2617-0561 |
| Aida Jafari | MSc | aida.Jafarey@gmail.com | https://orcid.org/0009-0003-3858-2886 |
| Parisa Rostami | MD | parisa.rostami.md@gmail.com | https://orcid.org/0009-0005-3336-0402 |
| Kais I. Abdul-Lateef Al-Abdullah | PhD | k.alabdullah@au.edu.kw | https://orcid.org/0000-0002-1315-9702 |
| Ryszard Tadeusiewicz | PhD | rtad@agh.edu.pl | https://orcid.org/0000-0002-4317-2801 |
| Paweł Pławiak | PhD | plawiak.pawel@gmail.com | https://orcid.org/0000-0001-9675-5819 |
| Roohallah Alizadehsani | PhD | roohallah.alizadehsani@qut.edu.au | https://orcid.org/0000-0003-0898-5054 |
| Siamak Pedrammehr | PhD | s.pedrammehr@tabriziau.ac.ir | https://orcid.org/0000-0002-2974-1801 |

## Abstract

Thyroid cancer is an increasing global health concern that requires advanced diagnostic methods. The application of AI and radiomics to thyroid cancer diagnosis is examined in this review. A review of multiple databases was conducted in compliance with PRISMA guidelines until October 2024. A combination of keywords led to the discovery of an English academic publication on thyroid cancer and related subjects. 368 papers were returned from the original search after 112 duplicates were removed. Relevant studies were selected according to predetermined criteria after 176 articles were eliminated based on an examination of their abstract and title. After the comprehensive analysis, an additional six studies were excluded. Among the 42 included studies, radiomics analysis, which incorporates ultrasound (US) images, demonstrated its effectiveness in diagnosing thyroid cancer. Various results were noted, some of the studies presenting new strategies that outperformed the status quo. The literature has emphasized various challenges faced by AI models, including interpretability issues, dataset constraints, and operator dependence. The synthesized findings of the 42 included studies mentioned the need for standardization efforts and prospective multicenter studies to address these concerns. Furthermore, approaches to overcome these obstacles were identified, such as advances in explainable AI technology and personalized medicine techniques. The review focuses on how AI and radiomics could transform the diagnosis and treatment of thyroid cancer. Despite challenges, future research on multidisciplinary cooper-ation, clinical applicability validation, and algorithm improvement holds the potential to improve patient outcomes and diagnostic precision in the treatment of thyroid cancer.

# 1. Introduction

Thyroid cancer ranks seventh globally in incidence with 820,000 cases in 2022, and 24th in mortality with approximately 47,000 deaths. Strikingly, women face a tripled risk of thyroid cancer. This highlights the necessary need for research, specifically to overcome radioiodine therapy resistance [1]. Of all endocrine neoplasms, thyroid carcinoma is the most frequent. Its incidence is continuously rising globally, ranking it as the fifth most frequent solid tumor in women and the sixteenth most common in males[1]. To effectively plan treatments and manage patients, an accurate diagnosis must be made quickly. Physicians mainly use imaging tests like computed tomography (CT), ultrasound (US), and magnetic resonance imaging (MRI) to figure out what kind of thyroid nodule you have. Experienced radiologists may find it challenging to interpret these imaging tests, though, as they must be able to identify subtle cancerous signs. Quantitative feature analysis of medical image data is the focus of the medical imaging subfield known as radiomics [2]. These features include textural, morphological, and intensity characteristics that are methodically extracted to disclose information that has been hidden in images obtained using modalities like computed CT, MRI, or positron emission tomography (PET) [3]. Using advanced computational techniques, radiomics aims to convert conventional medical images into data that can be analyzed. Radiomics enables the detailed characterization of tissues, tumors, and other anatomical structures by extracting a multitude of quantitative features [4]. Radiomics has a wide range of applications in healthcare, with a particular focus on situations involving cancer and is important for the identification, formulation, and assessment of cancer treatments. In order to analyze the extracted features, spot complex patterns, and possibly find biomarkers for prognostic and diagnostic uses, artificial intelligence (AI) and machine learning (ML) algorithms are essential [5].

Radiomics is particularly helpful in the field of tumor analysis when it comes to assessing the heterogeneity within tumors, forecasting treatment outcomes, and enabling personalized medicine by tailoring treatments based on distinct tumor characteristics [6]. The combination of intricate mathematical and computational methods with medical imaging not only improves our understanding of images particularly health-related ones but also presents opportunities for future advancements in the field. This progression demonstrates the continuous advancement and improvement of methods that link mathematical sciences, computational techniques, and medical imaging to enhance prognostic and diagnostic capabilities [7].

In the contemporary landscape of thyroid cancer diagnosis, there has been a discernible surge in research endeavors, particularly those delving into the integration of radiomics and AI. The heightened interest in this specific intersection of technologies reflects a growing recognition of their potential to reshape and enhance diagnostic approaches [8]. As the number of studies exploring the application of radiomics and AI in thyroid cancer diagnosis continues to multiply, there emerges an imperative to conduct a thorough and exhaustive review. This review aims not only to synthesize the current state of knowledge but also to identify gaps, challenges, and opportunities for further investigation. By systematically analyzing the existing body of literature, we seek to provide a comprehensive understanding of the progress made and the avenues yet to be explored, thereby laying the groundwork for continued advancements in the integration of radiomics and AI for thyroid cancer diagnosis.

In this review, recent advancements in the integration of radiomics and AI for thyroid cancer are explored, with a specific focus on: 1) A comprehensive examination of AI, ML, and DL in the context of thyroid cancer. 2) An in-depth overview of thyroid cancer, considering epidemiological and pathophysiological aspects. 3) An assessment of various techniques and methods employed in thyroid cancer diagnosis. 4) The development of radiomics and AI models in the field of radiomics. 5) Compilation and analysis of studies reporting outcomes of radiomics in thyroid cancer diagnosis through the application of AI techniques. The aim of this review is to provide valuable insights for researchers in the AI domain, aiding them in the creation and enhancement of efficient, high-performance tools for the diagnosis and treatment of thyroid cancer. Additionally, it is anticipated that this review will promote interdisciplinary collaboration, encouraging future experimental and theoretical work in healthcare, particularly in the realm of thyroid cancer diagnosis and treatment.

## 2. Literature search method

A thorough analysis of the body of research was conducted. This involved utilizing various scholarly scientific databases, including PubMed, Medline, EMBASE, Scopus, and Web of Sciences. The investigation focused on academic publications published up until October 1, 2024, and was restricted to those available in the English language. Additionally, supplementary articles were sourced by examining the reference sections of screened papers. An intricate search strategy was devised, incorporating a range of keyword combinations such as "radiomics," "artificial intelligence," "deep learning," "machine learning," "thyroid cancer," "thyroid neoplasm," and "thyroid tumor." These search strings were carefully constructed by combining keywords with their synonyms. Study selection involved initially identifying 368 research papers. Subsequently, 112 redundant documents were eliminated, and a review of titles and abstracts by authors [MY, SF, AiJ] resulted in the exclusion of 176 articles that did not meet predetermined criteria. A comprehensive review of full-text articles was then conducted by authors [AIJ, MY]. All review and conference studies were eliminated, leading to the exclusion of six additional studies. Subsequently, titles and abstracts were screened to select relevant studies based on inclusion and exclusion criteria, resulting in a final set of 42 studies [6, 9-50]. Figure 1 presents a detailed flow chart of the study selection procedure.

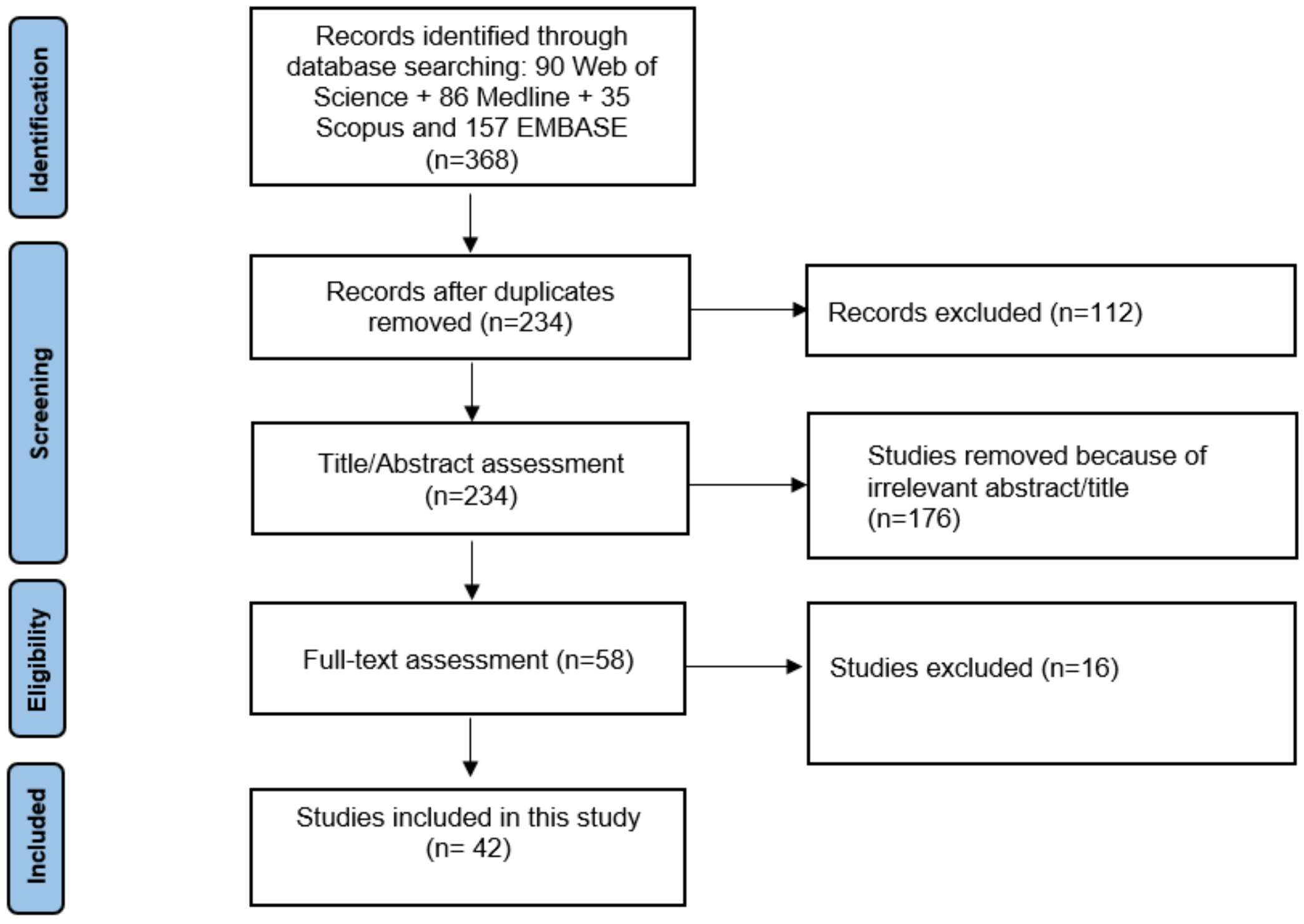

Figure 1 Review process flow diagram

## 3. Overview of Thyroid Cancers

### 3.1 Epidemiology

Based on the GLOBOCAN 2022 database of cancer incidence and mortality by the WHO International Agency for Research on Cancer, Thyroid cancer, a malignancy arising in the butterfly-shaped gland at our throat's base, may not rank highest in mortality, but its global reach demands attention. The year 2022 saw it claim the unfortunate title of seventh most common incidence (820,000 cases) of cancer (Figure 2) and 24th most common in mortality (about 47000 deaths), with over 83,000 diagnosed cases (*Figure 3*) [51]. While enhanced detection methods contribute to this number, a concerning shadow of overdiagnosis looms over specific regions, urging deeper understanding. This shadow falls unevenly, with women disproportionately bearing the burden. They are three times more likely to develop thyroid cancer, particularly the papillary and follicular types, which, thankfully, often boast favorable prognosis [52]

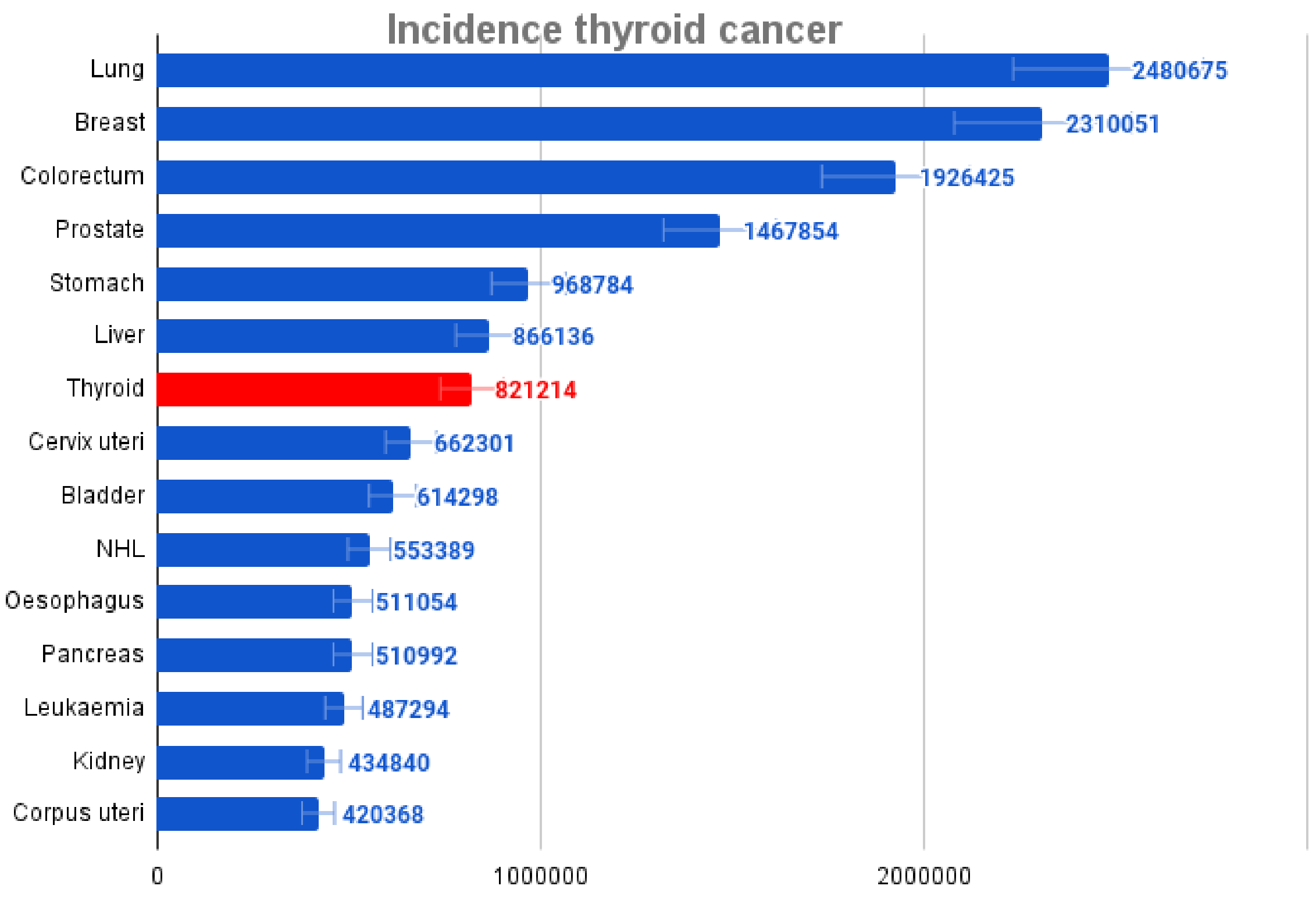

Figure 2 Presents the Incidence ranking of thyroid cancer among other types of cancers according to GLOBOCAN data from 2022.

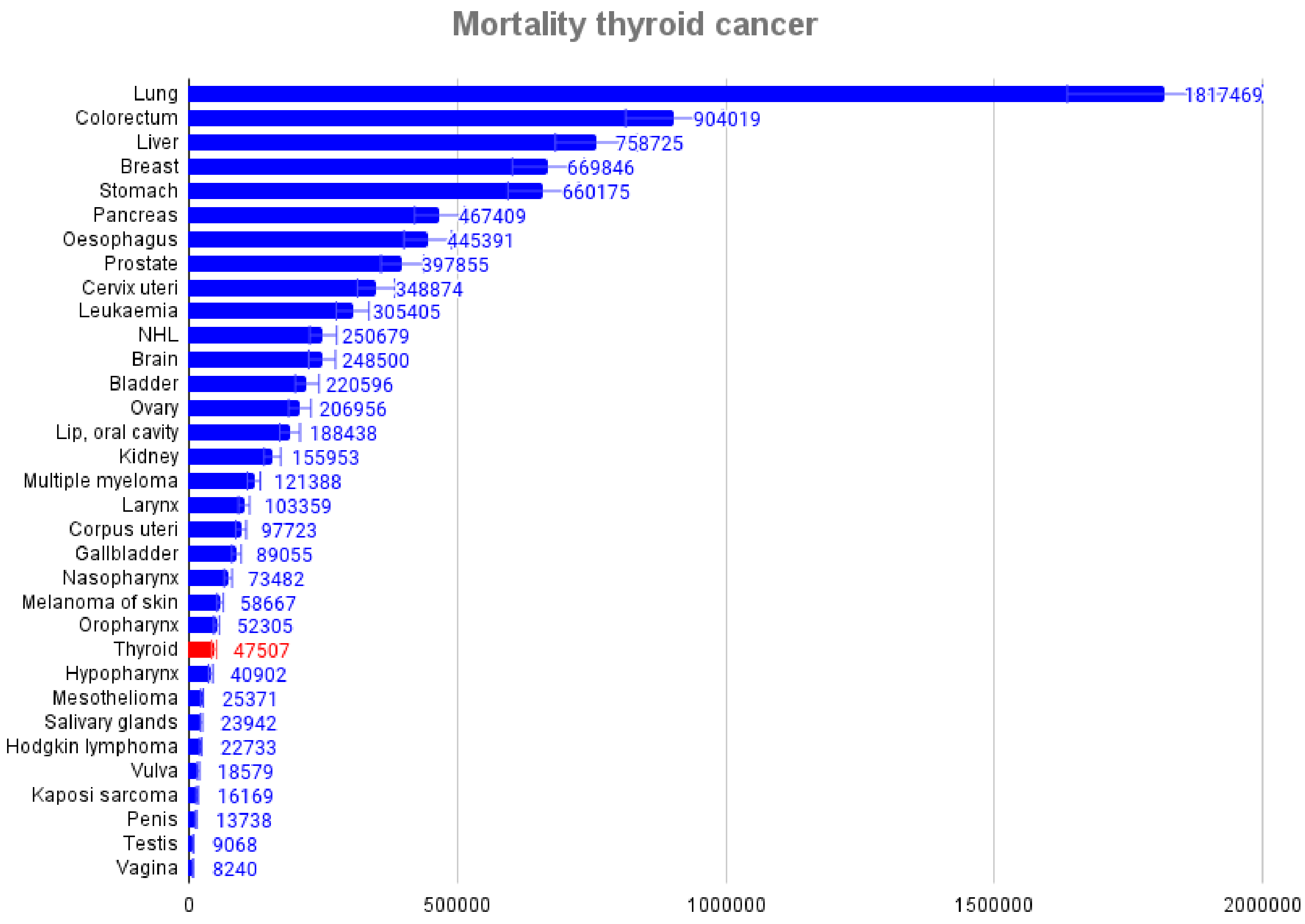

Figure 3 Presents the mortality ranking of thyroid cancer among other types of cancers according to GLOBOCAN data from 2022.

However, not all thyroid cancers share the same characteristics. Medullary cancer, although less prevalent, introduces a unique challenge with its slow growth and limited treatment options. On the opposite end of the spectrum, anaplastic carcinoma emerges as an aggressive entity with a nearly 100% mortality rate. This stark contrast in outcomes highlights the intricate nature of this disease [53].. Both preclinical and clinical settings are buzzing with activity, providing a ray of light for patients grappling with this formidable diagnosis. Beyond mere statistics, a deeper examination reveals that the median age of diagnosis typically falls in the early 50s [54]. However, this silent predator can strike at any stage of life, with the alarming distinction of being the most common malignancy among young adults aged 16-33. Unraveling the root causes remains crucial. Exposure to radiation, pre-existing thyroid conditions, environmental toxins, iodine deficiency, and even obesity all participate in the complex dance of risk factors. Understanding these intricate relationships empowers preventive measures and early detection efforts [55].

### 3.2 Pathophysiology

Thyroid abnormalities are divided into two benign and malignant groups. Nodular goiter, stemming from iodine deficiency, exhibits distinct classifications as endemic (>10% prevalence) or sporadic (≤10%). While endemic cases result from absolute iodine deficiency, sporadic instances involve various factors, such as increased thyroid hormone needs, dietary iodine insufficiency, goitrogen ingestion, and subclinical dyshormonogenesis [56]. Pathologically, nodular goiter manifests as a combination of hyperplastic and involutional changes, with challenges arising in distinguishing between follicular adenomas and hyperplastic nodules, particularly in the context of dominant nodules [57]. Beyond benign cases, encapsulated non-functioning nodules called follicular adenomas are common in people between the ages of 30 and 50 and can seldom result in hyperthyroidism [58]. A thorough histological examination is necessary to distinguish between adenoma and carcinoma. Women in their fifth decade make up the majority of patients with follicular carcinomas [59]. Malignant cells that are not fully differentiated exhibit behavior halfway between that of differentiated and undifferentiated carcinomas. They are characterized as solid masses with a variety of histological features [60]. Anaplastic carcinomas are extremely malignant and have a poor prognosis. They are composed of spindle and epithelioid cells. RET gene mutation screening is required for medullary carcinomas, which can be sporadic or familial. The former typically presents in the 40–60 age range, while the latter typically presents earlier and in both directions. In papillary carcinomas, BRAF and RET mutations activate the MAP kinase signaling pathway; in follicular lesions, PAX8/PPARγ and Ras mutations do the same. These molecular pathways are responsible for the differences between papillary carcinomas and follicular lesions [61]. Germline mutations in the RET proto-oncogene are linked to medullary carcinoma, especially in MEN2 syndromes [62].

### 3.3 Thyroid Cancer Diagnosis

Palpation for thyroid gland nodules was a major diagnostic tool in the past, but it is now only used in 30–40% of cases to identify thyroid cancers. A multimodal approach involving laboratory testing, imaging modalities, and clinical assessment forms the basis of modern thyroid cancer diagnosis [63]. The detection of thyroid cancer can now be achieved through a range of imaging modalities, such as neck and carotid USs, as well as scans of the chest, neck, and spine [64]. The ability to identify thyroid cancers even in the absence of palpable symptoms is improved by this extensive range of diagnostic techniques [27].

One of the most important steps in the diagnosis of recently formed thyroid nodules is measuring serum thyroid-stimulating hormone (TSH) levels to assist in differentiating between nonfunctional [65]. Hyperfunctioning nodules, which are rarely malignant, are indicated by subnormal TSH levels; non-functioning nodules or those with Hashimoto's thyroiditis may be more susceptible to cancer. Although a diagnostic neck ultrasonography cannot definitively determine whether a nodule is malignant, it is recommended to confirm the presence of nodules and assess any suspicious features. Fine needle aspiration (FNA) is necessary due to the elevated risk of malignancy, regardless of nodule size, especially for non-functioning nodules and if TSH is normal or elevated [66]. The outcomes of a FNA biopsy are categorized as benign, malignant, suspicious, indeterminate, or nondiagnostic concerning the diagnosis of thyroid cancer. [67]. Distinct histological variants of thyroid cancer, including anaplastic, follicular, medullary, and papillary (*Figure 4*) [67, 68].

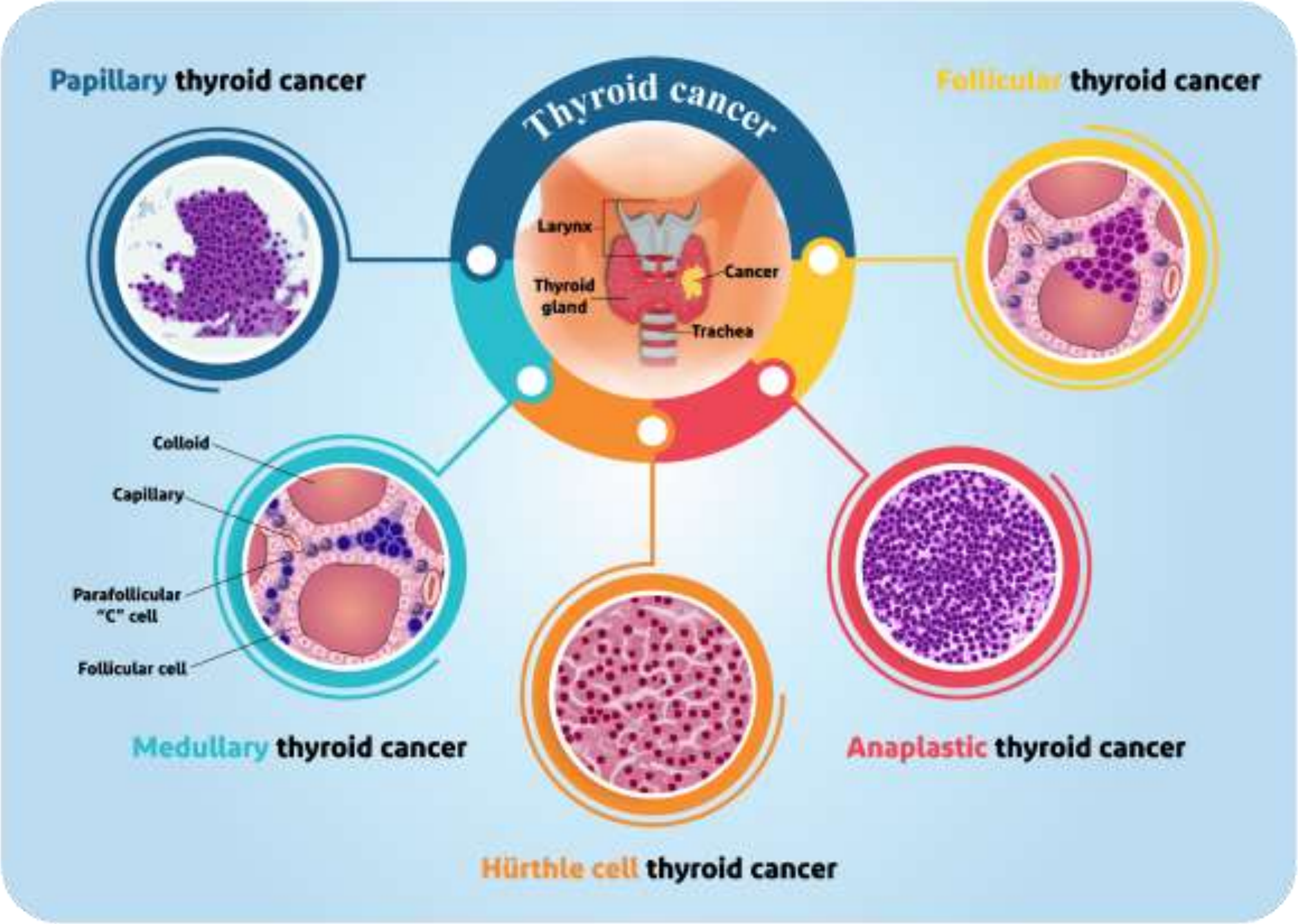

Figure 4 There are different types of thyroid cancer including:  papillary thyroid cancer, follicular thyroid cancer, oncocytic cell (hurthle cell) thyroid cancer, medullary thyroid cancer, anaplastic thyroid cancer

The realm of thyroid cancer diagnosis and treatment has experienced enhanced efficacy with the escalating application of radiomics in recent years (*Figure 5*) [67, 68].

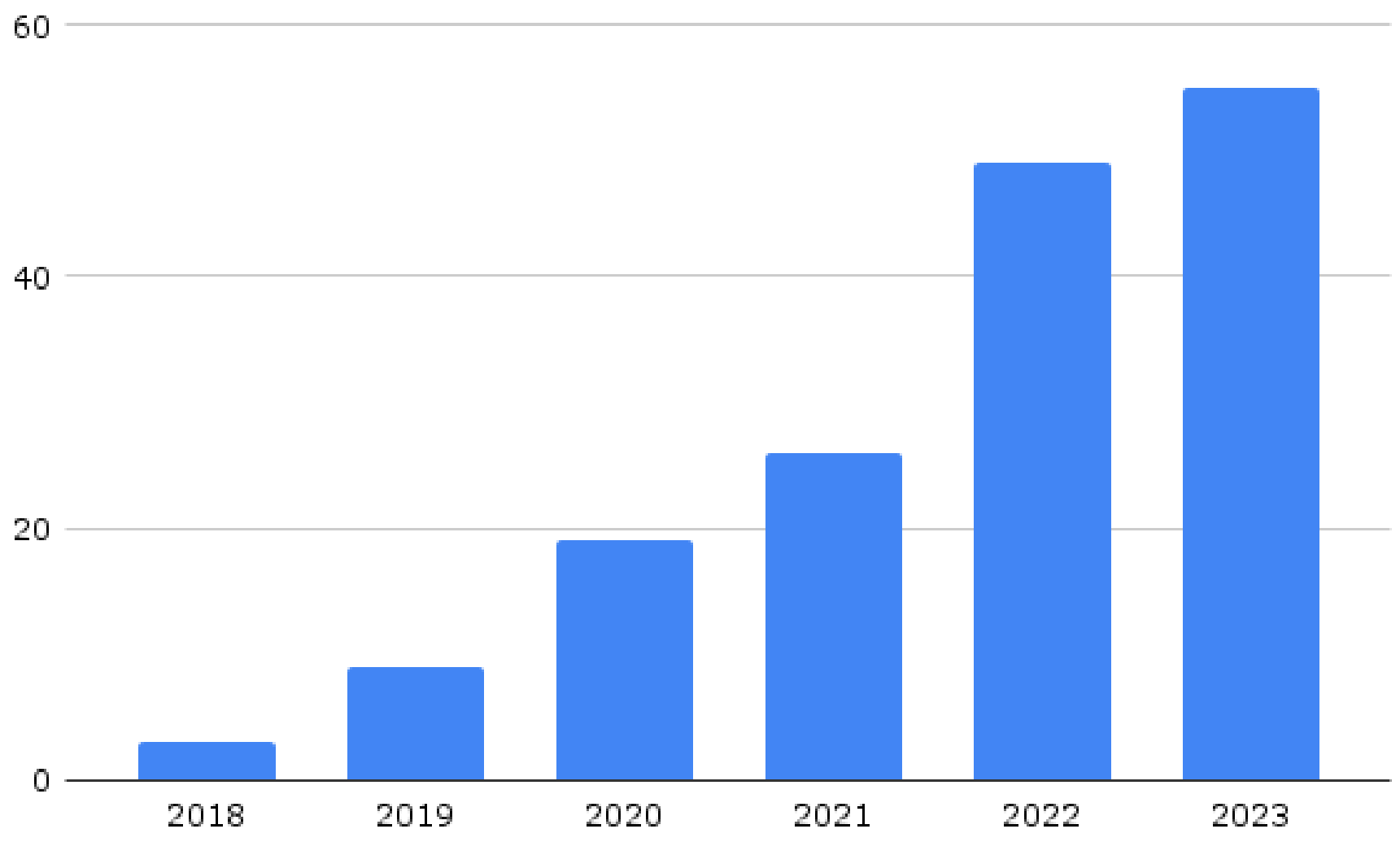

Figure 5 A bar chart display increasing investigations over the last 6 years about utilizing radiomics and AI in Thyroid cancer diagnosis

## 4. Overview of artificial intelligence

The goal of AI, a fast-developing area of computer science, is to create machines that possess traits of human intelligence. AI encompasses a wide range of methodologies, but some of the most well-known ones are natural language processing (NLP), DL, and ML [69]. Notably, Large Language Models (LLMs) constitute a unique class in the AI algorithmic framework that is capable of understanding, compressing, producing, and forecasting textual content by employing large-scale datasets and advanced DL methods. Given their inherent capacity for text generation, LLMs find versatile applications in various NLP tasks, including sentiment analysis, translation, summarization, rewriting, and classification [70].

A branch of AI called, NLP studies how human language is conceived, decoded, and understood. This broad field includes a variety of approaches, including sentiment analysis, speech recognition, machine translation, and text mining. AI's historical trajectory has undergone notable shifts, moving from the first age of rule-based systems to the current era where ML and DL algorithms are the mainstays (Figure 6) [71-73].

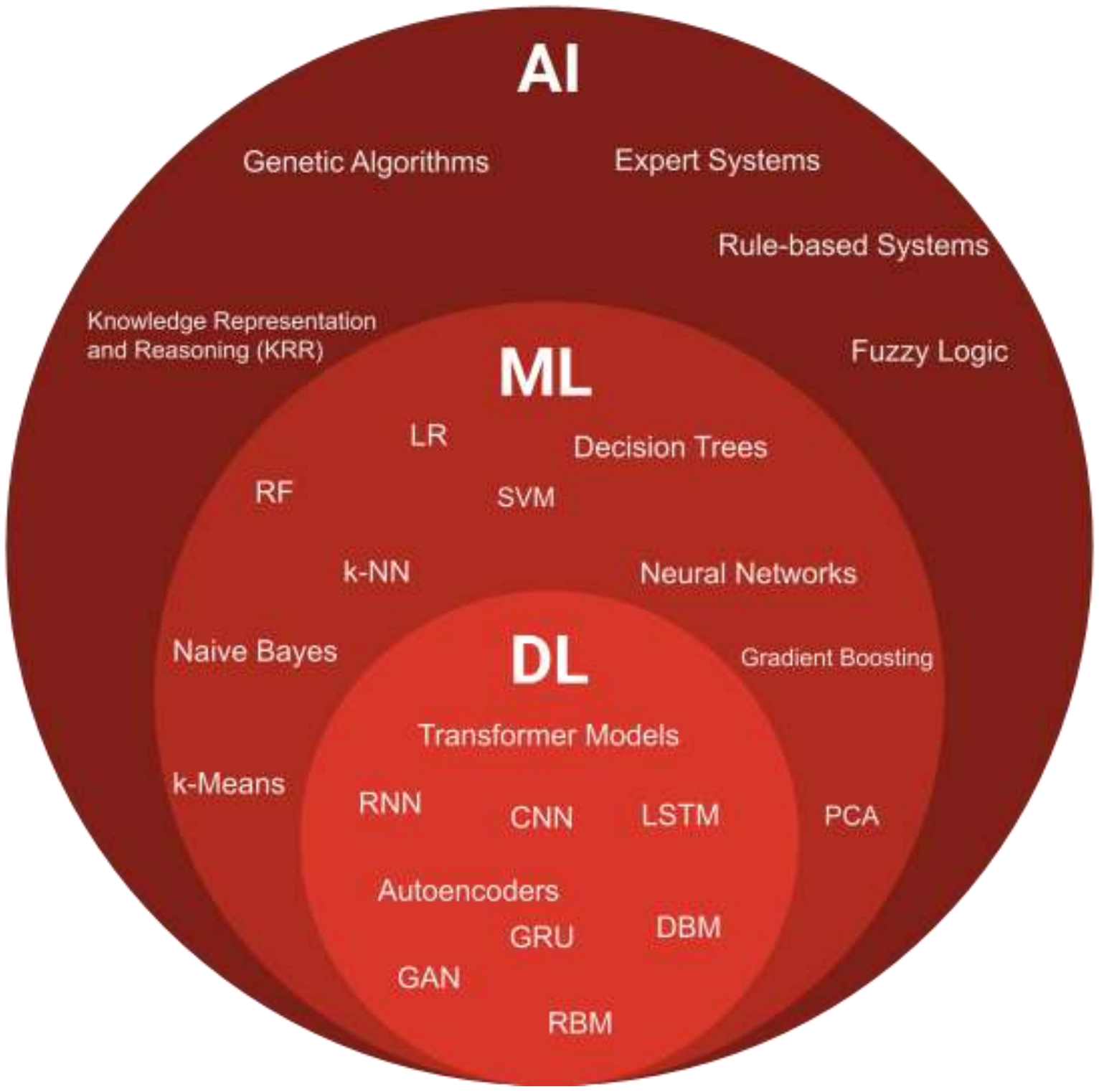

Figure 6 In ML algorithms: Linear Regression (LR), Random Forest (RF), Support Vector Machines (SVM) k-nearest neighbors (k-NN), Principal Component Analysis (PCA), Gradient Boosting Algorithms (e.g., XGBoost, LightGBM). And in DL algorithms: Convolutional Neural Networks (CNN), Recurrent Neural Networks (RNN), Long Short-Term Memory (LSTM), Gated Recurrent Unit (GRU), Generative Adversarial Networks (GAN), Deep Boltzmann Machines (DBM), Restricted Boltzmann Machines (RBM), Deep Q-Networks (DQN)

## 4.1 Machine learning

Within the field of AI, ML allows computers to grow without explicit programming by using experience as a learning tool. The primary objective is to develop algorithms that enable computers to identify patterns in data and extrapolate inferences or predictions [74]. A large amount of data (referred to as the training dataset) is fed into the system as part of standard ML procedures, and the algorithm is then allowed to search for patterns, relationships, and features in the data. By adjusting its internal parameters, the system gains experience and improves its ability to complete tasks. Artificial neural networks (ANNs), which resemble learning systems, are used in DL [75]. ML can be divided into three main categories (*Figure 7*):

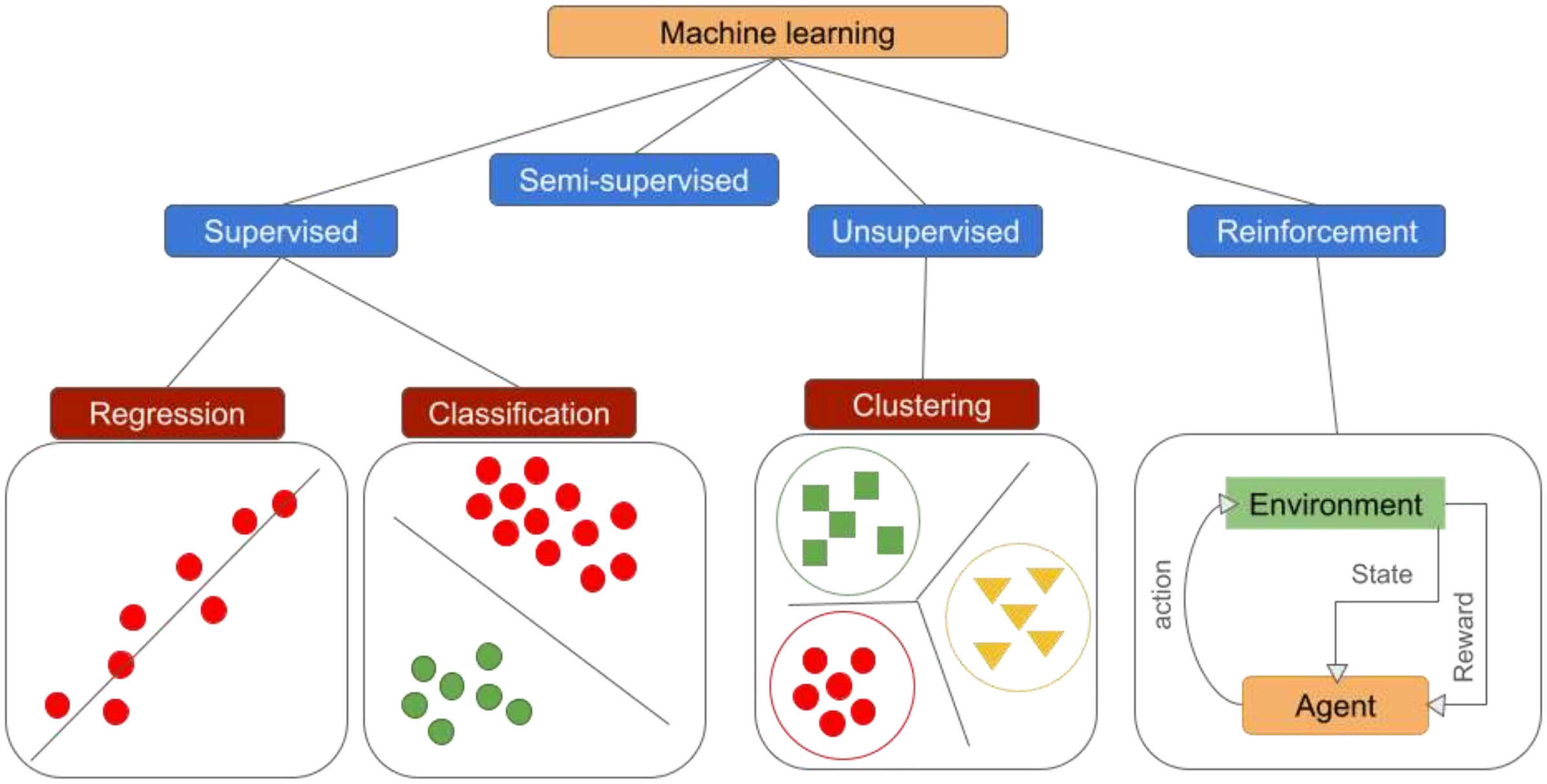

Figure 7 The diagram depicts three learning paradigms: supervised, unsupervised, and reinforcement learning. Within the supervised category, the focus is on predicting outcomes through regression and classification. In the unsupervised realm, the emphasis is on revealing patterns by employing clustering techniques. Lastly, the reinforcement learning section illustrates the dynamic process of decision-making, showcasing actions, states, and rewards.

**Supervised Learning:** In this approach, the algorithm is trained on a labeled dataset, where the input data is paired with the corresponding desired outputs. The goal is for the algorithm to learn a mapping function that can predict the output for new, unseen inputs [76, 77].

**Semi-supervised**: Semi-supervised learning occupies a middle ground between supervised and unsupervised learning methodologies within the realm of ML. This approach utilizes a limited set of labeled data in conjunction with a vast pool of unlabeled data for training purposes [78]. This hybrid training strategy proves advantageous when only a portion of the data has ground-truth labels available. Moreover, the amalgamation of labeled and unlabeled data has been observed to enhance the accuracy of classifying new instances. Examples of semi-supervised learning algorithms include restricted Boltzmann machines and deep belief networks [79].

**Unsupervised Learning:** In this case, the algorithm is entrusted with identifying patterns or structures in the unlabeled data without the need for explicit guidance. Unsupervised learning frequently involves tasks like clustering and dimensionality reduction [80].

**Reinforcement learning**: Systems can learn by interacting with real-world environments thanks to a subset of ML called reinforcement learning. The computer attempts to solve problems by making mistakes and trying different things. An entity referred to as the agent engages with the environment during this process, making observations, choosing actions, and getting rewarded or penalized in accordance with those decisions [81]. The agent's main goal is to maximize cumulative rewards; the best course of action is known as the policy. Numerous fields, such as game theory, data science, and robotics, use reinforcement learning [82].

Applications of ML are found in many different fields, such as recommendation systems, fraud detection, NLP, image and speech recognition, and autonomous cars. As technology develops, ML models and DL models in particular become more and more popular because of their capacity to automatically extract hierarchical features from data, which qualifies them for difficult tasks [83].

## 4.2 Deep learning

As mentioned, fueled by advancements in computer processing, a subset of ML rooted in ANNs has emerged to address progressively intricate challenges autonomously. While early ANNs necessitated substantial human input for feature selection, the advent of DL applications has revolutionized this process by automatically extracting optimal features from raw data [84]. DL is predominantly deployed in pattern recognition systems, showcasing its efficacy in extracting crucial information from high-dimensional datasets. Notably, DL methods have exhibited promising outcomes, particularly in computer vision and image analysis, often rivaling or surpassing traditional ML approaches reliant on manually crafted features defined by human experts [85]. DL can autonomously carry out a designated task by extracting crucial features directly from unprocessed data. A DL method is characterized by its composition of multiple layers, forming a hierarchical structure of simple models. These models nonlinearly transform their input to generate output, creating a complex mapping process [86]. Each layer in this stack further refines the outputs from the previous layer, contributing to the enhancement of both selectivity and invariance within the representation. This iterative refinement process allows DL models to capture increasingly intricate patterns and features in the data. Overall, the multilayered architecture enables DL systems to achieve high levels of performance in various tasks, ranging from image recognition to NLP [87]. As an instance, Peng et al. [88] introduced ThyNet, an AI model based on DL, to aid in the diagnosis and treatment of thyroid nodules. They used a dataset of 18,049 US images and involves 8,339 patients. The purpose of ThyNet is to differentiate between benign and malignant thyroid nodules in order to improve diagnostic precision and minimize needless FNAs. In this study, radiologists use ThyNet to assist in the diagnostic process. ThyNet achieved a significantly higher AUROC (0.922) than radiologists (0.839). ThyNet assistance improved radiologists' AUROC from 0.837 to 0.875 in image review and from 0.862 to 0.873 in clinical tests. In simulations, ThyNet reduced FNA from 61.9% to 35.2% and lowered missed malignancy rates from 18.9% to 17.0%. According to the study, ThyNet had much better accuracy and AUROC than radiologists alone. Both junior and senior radiologists performed better diagnoses with ThyNet's help. ThyNet also showed promise in lowering needless FNAs, which may have an effect on clinical decision-making. ThyNet's efficacy in the real-world clinical setting is attributed to the particular DL architecture and methodology employed, including the training process on a large dataset and the integration of ThyNet with the ACR TIRADS classification.

## 5. Radiomics

In the world of biology and medicine, many fields end in "-omics" to signify the analysis of massive data sets to uncover hidden insights. Radiomics is one such field, specifically focused on extracting valuable, measurable information from medical images, helping us unlock secrets hidden within US images, CT, MRI, or PET/CT [4]. Using complex algorithms, radiomics analyzes medical images to extract a multitude of "radiomic features". Imagine being able to extract hidden information from a medical image, rather than just seeing shapes and shadows. That's what radiomics does [89]. It uses complex algorithms to mine medical images for a vast number of tiny details, revealing patterns and characteristics invisible to the naked eye.

This approach unlocks hidden insights about tumors, potentially predicting how they behave and respond to treatment [90]. This emerging field, born from radiology and oncology, holds immense promise for personalized medicine, tailoring treatments based on the unique characteristics of each patient's disease [91]. While currently most advanced in these fields, its potential extends to any area where imaging unveils clues about disease processes. In recent years, a discernible integration of applied mathematics and statistical sciences has been witnessed across diverse domains, particularly in the development and optimization of AI and ML algorithms [92]. This interdisciplinary approach, heavily anchored in computer science and computational methods, has fortified the underpinnings of AI. Matrices, employed in image processing, have garnered significance, propelling the evolution of CNNs. This evolution has significantly altered the landscape of image analysis, especially within fields such as medical imaging. The utilization of matrices in image description has concurrently given rise to innovative techniques, including radiomics [91].

### 5.1. Data acquisition

Image acquisition, the first crucial step in radiomics, acts as the gateway to unveiling hidden information within the patient. This intricate process involves capturing medical images that serve as the foundation for subsequent feature extraction and analysis [93]. Beyond CT, PET, and MRI, radiomics embraces US's unique strengths and challenges. Its real-time visualization, lack of radiation, and cost-effectiveness make it ideal for dynamic assessments, monitoring, and point-of-care applications [94]. However, operator dependence, limited penetration, and speckle noise demand careful standardization, advanced techniques like elastography and ML, and tailored protocols. Despite these hurdles, US is unlocking its potential in radiomics for early cancer detection, assessing aggressiveness, monitoring treatment, and more, solidifying its role as a transformative force in personalized medicine [95].

### 5.2. Region of Interest (ROI) Segmentation

Within the burgeoning field of radiomics, the meticulous process of ROI segmentation occupies a pivotal role, particularly when unraveling the complexities of thyroid cancers. Akin to a sculptor carefully excavating a hidden masterpiece, this technique meticulously delineates the tumor region within medical images, enabling the extraction of quantitative features crucial for diagnosis, prognosis, and treatment stratification. Let us delve deeper to illuminate the significance of ROI segmentation in this context [4, 96].

**5.3 Fiduciary Accuracy: The Foundation of Feature Extraction**

Envision setting out on a scientific quest to find valuable features that will reveal the mysteries of tumor biology, rather than buried treasure. Accurate ROI segmentation acts as your steadfast guide, ensuring that the features that are extracted accurately capture the distinct qualities of the tumor, including its geometric structure, textural heterogeneity, and intensity distribution [97]. This precision is critical to building strong radiomics models, which are complex algorithms that analyze these features in-depth to distinguish between benign and malignant nodules, forecast aggressive behavior, and help choose the best course of treatment. Think of them as experienced detectives. Similar to a malfunctioning compass, inaccurate segmentation can result in dangerous misinterpretations that could misclassify a benign nodule as malignant or vice versa, with serious consequences for patient care [98].

**5.4 Converging on Standardized Segmentation to Close the Interobserver Gap**

Consider a scenario in which two detectives are looking into the same crime scene, but they each have a different perspective on it. Radiologists may exhibit intrinsic subjectivity and interobserver variability when manually segmenting a tumor by visually tracing its outline; for instance, one radiologist may perceive a jagged boundary while another sees a smoother one [99]. This discrepancy can make it extremely difficult to create trustworthy radiomics models. Automated and semi-automated techniques are becoming valuable allies in closing this gap. These techniques make use of advanced image processing algorithms to accurately and consistently identify the tumor, even when viewed by multiple observers and research projects. Envision a group of robotic investigators, outfitted with state-of-the-art equipment and unwavering accuracy, carefully examining the scene and removing any subjectivity that might skew human judgment [62, 96, 100].

**5.5 Revealing Latent Heterogeneity: The Prospects for Personalized Medicine**

Consider a tumor as a varied terrain with hidden subregions that each have unique biological characteristics, rather than as a single, monolithic structure. Sophisticated segmentation techniques that combine data from various imaging modalities, such as CT, PET, and US scans, are like making an intricate topographical map of this complex terrain. With unmatched granularity, these "multi-parametric" models are able to depict the tumor's heterogeneity and identify biologically distinct subregions. With this newfound understanding, a new era of personalized medicine may well be ushered in [101]. Medical practitioners may be able to tailor treatment regimens to target specific tumor regions by identifying these subregions, offering a more precise and effective way to treat the condition.

**5.6 Pushing the Boundaries: Seeking Unrivaled Accuracy**

The constant search for advancement is essential to the spirit of scientific inquiry. In order to improve ROI segmentation techniques, researchers are constantly looking for new approaches, especially in difficult situations where the tumor boundaries are hazy or overlap with adjacent structures [63]. To get even more accuracy, advanced image processing methods and DL algorithms are being used. Envision a novel type of detective, armed with state-of-the-art equipment and unparalleled analytical abilities, meticulously inspecting the crime scene and revealing every detail in their pursuit of the truth [99, 102].

**5.7 Working Together and Using Open-Source Tools to Create a Shared Future**

In addition to individual genius, the scientific community as a whole must contribute to the search for the best segmentation techniques. To create open-source software tools and standardize segmentation protocols, researchers are actively working together. Because of the increased accessibility and reproducibility, researchers from all over the world can use these resources and add to the body of knowledge. Envision an international community of investigators, exchanging their knowledge, perspectives, and resources to advance their respective specialties and, in the end, enhance patient results [103].

The foundation of radiomics-based analysis for thyroid cancers is ROI segmentation. It enables researchers to extract useful features, build reliable models, and open the door to personalized medicine by precisely defining the tumor [104]. Future research is expected to yield even more advanced segmentation strategies, which will fully realize the potential of radiomics to transform the treatment of thyroid cancer. And who knows? Maybe these developments will motivate a new wave of medical sleuths to use knowledge and technology to solve medical mysteries and save countless patients' lives [103, 105].

**5.8 Extraction of features**

In radiomics, feature extraction is essential for revealing patterns and information from medical images that are not immediately visible to the human eye. This is particularly true when analyzing thyroid cancer. To assist with diagnosis, prognosis, and treatment planning, feature extraction extracts different aspects of the images using complex algorithms and computational tools [106]. This process involves identifying and quantifying a wide range of meaningful characteristics from the image data in order to obtain significant insights into the underlying tissue characteristics and potential abnormality indicators within the thyroid gland (*Figure 8*) [107].

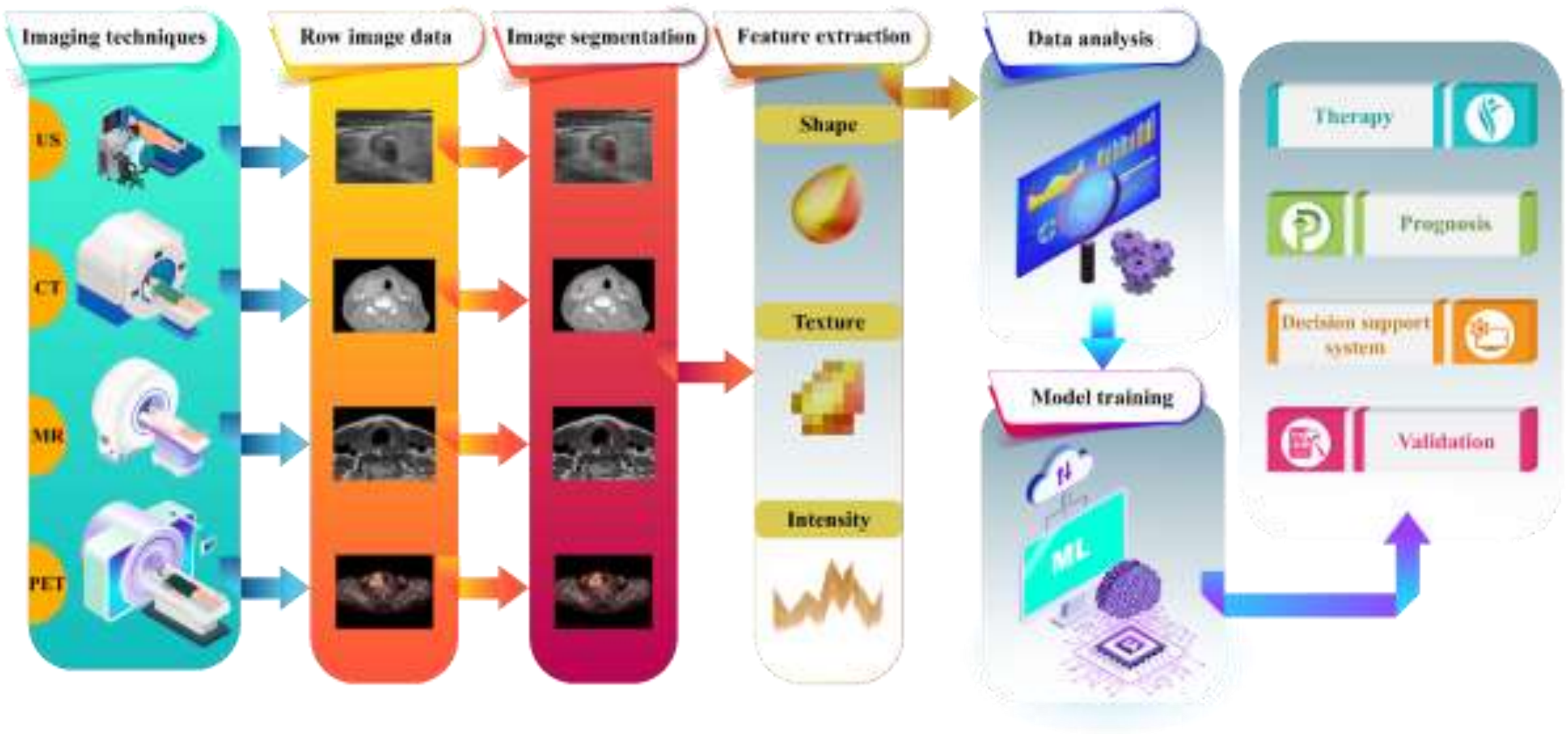

Figure 8 Radiomics analysis involves: (1) Image Acquisition, (2) Image Preprocessing, (3) ROI Segmentation, (4) Feature Extraction, (5) Feature Selection, (6) Model Development, (7) Validation, (8) Interpretation and Reporting, and (9) Clinical Translation.

### 5.9 Features of shape

These characteristics include geometric attributes like the thyroid gland's size, area, roundness, and eccentricity of its regions. Researchers and radiologists can learn more about the spatial arrangement and structural traits of tissues, which may be suggestive of malignant growth or other anomalies, by examining the shape descriptors [103].

### 5.10 Features of intensity

The distribution of pixel values throughout the image is the main focus of intensity features. By extracting information about variations in tissue density or contrast within the thyroid region, analysis of intensity histograms can help identify lesions that may be cancerous [105, 107-109].

### 5.11 Textural features

Statistical metrics like GLCM and GLRLM, which reveal spatial correlations between pixel intensities, are used in texture features. By evaluating texture features, radiomics can help differentiate between benign and malignant thyroid nodules by providing insight into tissue heterogeneity, smoothness, and coarseness [110].

### 5.12 Wavelet Features

By breaking down the image data into distinct frequency components, wavelet analysis makes it possible to extract spatial-frequency information. Wavelet features aid in the capture of texture patterns at different scales, providing a comprehensive analysis of the composition of thyroid tissue and possibly exposing minute alterations in texture linked to the development of thyroid cancer [108, 109].

In order to extract minute details from medical images, feature extraction in radiomics for thyroid cancer analysis entails a careful examination of shape, intensity, texture, and wavelet features. These characteristics act as quantitative biomarkers to help doctors make well-informed decisions regarding patient care, diagnosis, and treatment planning. Through the use of advanced computational methods, radiomics plays a significant role in providing personalized and data-driven approaches to thyroid cancer assessment, ultimately contributing to precision medicine initiatives [108-110].

### 5.13 Feature selection and feature dimension reduction

Feature selection, which comes after radiomic feature extraction in radiomics, is a crucial step in enhancing the effectiveness of ML algorithms for thyroid cancer data analysis. linked to high dimensionality, enabling quicker model training, and improving the repeatability of the acquired outcomes. Feature selection serves as a filtering mechanism to concentrate on the most pertinent characteristics that greatly aid in differentiating between various thyroid cancer phenotypes and assisting in accurate diagnostic outcomes. It does this by choosing the most informative features from the pool of extracted characteristics [111].

To optimize the selection process in the particular context of thyroid cancer analysis, a thorough investigation included six distinct feature selection techniques. ANOVA, forward and backward elimination, Pearson correlation, ridge regression, and the least absolute shrinkage and selection operator (LASSO) were some of these techniques. By employing diverse feature selection methodologies, the investigators sought to capitalize on the advantages of each approach in pinpointing the most pivotal radiomic features that encapsulate critical details pertaining to tumor attributes, thereby facilitating a more precise and streamlined examination of thyroid cancer imaging data [112].

The three primary categories of feature selection techniques used in this study are filter, wrapper, and embedded techniques. By using statistical tests to assess the relationship between features and tumor aggression, filters like Pearson correlation and ANOVA choose features according to their relative importance. By repeatedly modifying thresholds, wrapper techniques such as forward and backward elimination construct subsets of features with the goal of assembling a set of variables with strong predictive ability. Feature selection is integrated into the model training process through embedded techniques like LASSO and ridge regression, which combine elements of both filter and wrapper methods to balance model complexity and interpretability while optimizing predictive performance [112, 113].

The application of filter, wrapper, and embedded techniques in feature selection denotes a multimodal strategy for determining the most pertinent radiomic features for analysis of thyroid cancer. While wrapper methods iteratively refine feature subsets to improve prediction accuracy, filter methods help with the initial screening of features based on their statistical significance [107]. In the context of thyroid cancer radiomics, embedded techniques seamlessly integrate feature selection with model training, utilizing the advantages of both filter and wrapper methods to achieve a balance between feature relevance, model interpretability, and predictive performance. Medical researchers are using DL more and more to analyze medical images. By using this method, they can directly collect a greater amount of data rather than having to go through the laborious process of manually extracting features from the images. In recent years, this has gained popularity as a trend [114, 115].

**5.14 Model development**

Radiomics utilizes advanced image analysis techniques to extract valuable information hidden within medical scans. This information, known as radiomic features, goes beyond the visual capabilities of traditional diagnosis. These features capture subtle characteristics of tissues, like their structure, texture, and intensity, similar to a unique fingerprint. By combining these features with powerful modeling techniques and extensive databases, Radiomics empowers the development of classifiers. These classifiers, like sophisticated detectives, analyze new medical images and predict potential outcomes, ultimately shaping the field of personalized medicine [105].

Currently, Radiomics employs a diverse toolbox of ML algorithms for building these classifiers [116]. This toolbox includes established methods like LR, RF, SVM, and DT. Additionally, innovative techniques like k-nearest neighbors, ANNs, and Bayesian and clustering algorithms are increasingly being explored. It's important to note that while most of these algorithms fall under the category of supervised learning, requiring pre-labeled data to train the models, some, like clustering algorithms, operate under unsupervised learning, uncovering hidden patterns without prior labels [117].

Building a robust classifier model involves a two-stage process: training and testing [118]. During the training phase, the model "learns" to distinguish between different outcomes based on the provided data. This data set, crucial for successful learning, needs to be large and representative of the broader population being studied. Once trained, the model is put to the test in the testing phase. Here, entirely new and unseen data is used to evaluate the model's accuracy and ability to accurately predict outcomes. When possible, a third stage, known as validation, can further enhance the model's performance. In this phase, the model parameters such as the quantity and significance of the various features employed are adjusted in light of the validation data [118]. Through careful navigation of these phases, Radiomics opens up new possibilities for medical images and ushers in a new era of customized healthcare.

## 6. Radiomics and AI in Thyroid Cancer Diagnosis

The use of radiomics and AI models to improve thyroid cancer diagnosis and treatment has increased significantly in the last few years. Numerous ML and DL techniques have been adopted in this rapidly developing field, showing encouraging results. Numerous studies have examined the effectiveness of these approaches in assessing ML and DL models for the diagnosis and treatment of thyroid cancer, with varying degrees of accuracy [119]. These efforts represent a noteworthy progression in utilizing computational methods to enhance the comprehension and treatment of thyroid cancer, with great promise for enhancing oncology patient outcomes and clinical procedures [120]. A selection of these studies is presented below based on the imaging type. These studies pave the way for future developments in precision medicine by rigorously evaluating and comparing data, which helps to validate and improve computational approaches for the diagnosis and treatment of thyroid cancer [121].

Among the selected studies (Table 1**Error! Reference source not found.**), the utilization of MRI images was relatively limited, with only three studies incorporating this modality into their analyses. MRI offers distinct advantages in capturing detailed anatomical information and soft tissue contrast, making it a valuable imaging tool in oncology. However, its application in thyroid cancer diagnosis and treatment remains less explored compared to other modalities such as ultrasound or CT [9]. Despite this, the studies that did incorporate MRI imaging demonstrated the potential for MRI-based radiomics to contribute to the understanding and management of thyroid cancer.

### 6.1. Radiomics and AI using MRI Imaging for Thyroid Cancer

A study by Wei et al.[10] used radiomics based on multiparametric MRI to predict extrathyroidal extension (ETE) in PTC. Thirty-two patients made up the test cohort and ninety-two patients made up the training cohort in their study. They created a radiomics predictive model that demonstrated the ability of MRI radiomics to stratify PTC patients based on ETE prior to surgery, achieving high AUCs of 0.96 and 0.87 in the training and testing sets, respectively. The study used a combination of high-performing features from DWI, T2 WI, and CE-T1WI images in a retrospective trial that involved manual segmentation of thyroid tumors for radiomic feature extraction. According to the findings, MRI-based radiomics may be able to predict ETE in PTC in a non-invasive manner, which would be helpful for preoperative evaluation and treatment planning.By combining features from contrast-enhanced T1-weighted, T2-weighted, and diffusion-weighted images, Dai et al.[11] developed a multimodality MRI-based radiomics model for predicting aggressiveness in PTC, achieving a high accuracy of 0.917. Their research, which included 107 PTC patients, attempted to provide guidance on prognosis and treatment options for severe cases. The study illustrated the potential of MRI-based radiomics in non-invasively assessing thyroid cancer aggressiveness by using a sparse representation method for feature selection and model establishment. Manual segmentation was performed for region of interest delineation, and the study highlighted the possibility of exploring DL for automatic segmentation in future work. The model's high accuracy, supported by an AUC of 0.960, underscores the clinical relevance and utility of ML-based image analysis in rapid and informative cancer assessment. H.Qin et al.[9], in a retrospective investigation examined preoperative MRI images from 109 PTC patients (77 with LN metastases and 32 without). They extracted radiomics signatures from fat-suppressed T2-weighted MRI images, identifying optimal features using Spearman correlation, hypothesis testing, and RF methods. The AUC for LN status based on visual diagnosis was 0.739 (sensitivity = 0.571, specificity = 0.906). The MRI radiomics model with five optimal signatures achieved AUCs of 0.805 in the training group and 0.760 in the validation group. The combined model of the best radiomics features and visual MRI assessment showed a high AUC of 0.969 (sensitivity = 0.938, specificity = 1.000), demonstrating superior diagnostic efficiency for LN metastases in PTC.

All three-research used different models and approaches to analyze and predict the progression of PTC. Specifically, the combined MRI radiomics model in Qin's [9] study provided better accuracy and specificity. these studies not only confirm the value of MRI-based radiomics in PTC, but also underline the importance of implementing personalized techniques that take into account individual clinical scenarios. This diversity of study ideas reflects radiomics' rising acceptance as a valuable tool in personalized medicine, with the ultimate goal of improving patient care and treatment outcomes in thyroid cancer.

### 6.2. Radiomics and AI using CT-Scan Imaging for Thyroid Cancer

Eleven studies employed CT images as part of their radiomics analyses in thyroid cancer research. CT imaging offers distinct advantages, including high spatial resolution and the ability to visualize anatomical structures in detail. This modality provides valuable information on tumor size, shape, and density, facilitating accurate diagnosis and staging of thyroid cancer. In these studies, CT-based radiomics approaches were utilized to extract quantitative imaging features from CT scans, enabling the development of predictive models for various aspects of thyroid cancer, such as lymph node metastasis, extrathyroidal extension, and cervical lymph node metastasis.

A CT-based radiomics model was created in a multicenter study directed by Li et al. [16] to forecast central cervical lymph node metastases (CLNM) in patients with PTC. By incorporating radiomic characteristics and clinical risk factors, the model showed superior sensitivity and accuracy compared to a skilled radiologist, indicating its potential as a non-invasive preoperative instrument with practical applications. The model, which focused on the linear-SVM algorithm and included 4227 extracted radiomic features, showed encouraging results and good agreement when predicting CLNM across a range of test sets. The linear-SVM algorithm achieved the highest AUC of 0.709 in the internal test set, outperforming other models (KNN, DT, LR, Gaussian-SVM) with AUCs of 0.605. It also surpassed CT assessments by experienced radiologists in both sensitivity (0.702 vs. 0.477) and accuracy (0.670 vs. 0.642). The combined radiomics model using linear-SVM showed AUCs of 0.747 in the training set, and 0.764 in the external test set.

The use of a CT-based radiomics model for predicting thyroid capsule invasion (TCI) in PTC was examined in a two-center study directed by Wu et al. [18] CT scans were processed to extract radiomic features, and six supervised ML models were built. In predicting TCI, the combined-linear-SVM model performed better than the radiomics and clinical models because of its stable and clinically useful integration of optimal radiomics features and clinico-radiological risk factors. In the internal test cohort, the combined-L-SVM model achieved the highest AUC of 0.820, outperforming both the best radiomics model (AUC = 0.733) and the clinical model (AUC = 0.709). This combined model integrates 23 radiomics features with 1 clinico-radiological risk factor (CT-reported TCI). In the external test cohort, it showed stability with an AUC of 0.776. Notwithstanding the limitations of the study, such as the absence of a sizable sample size and the consideration of various factors, the ML-based methodology exhibited strong predictive capacity.

Yu and colleagues [19] created a radiomics nomogram that combines clinical risk factors with radiomics features taken from CT images to predict ETE in PTC patients. The AUCs ranged from 0.642 to 0.701, indicating the potential value of the nomogram in predicting ETE, although the study was retrospective in nature. Certain limitations are recognized, such as the possibility of selection bias and the requirement for prospective studies to account for confounding variables. The study sought to address the effects of ETE, which are linked to lymph node metastasis and invade surrounding structures, both of which have an impact on outcomes. Using a variety of ML techniques, the radiomics nomogram yielded potential to support preoperative ETE prediction in PTC. 153 patients were included in this study; Four radiomics features were selected, and six models demonstrated the value of radiomics with AUCs ranging from 0.642 to 0.701. The nomogram combining the Rad-score and clinical risk factors (radiologists' interpretation) performed well, achieving AUCs of 0.750 in the internal test set and 0.797 in the external test set. .

A LR model is used in the study by Xu et al. [20] to examine the diagnostic potential of CT-based radiomics for differentiating between benign and malignant thyroid nodules. The research, which examined data from 161 patients, shows how accurate the model is, with an amazing AUC of 94.4% in the training group and 94.2% in the test group. The model also has corresponding sensitivities of 89.7% and 89.4% and specificities of 86.7% and 86.8%. The study's limitations, including its small sample size and dependence on single-center data, are acknowledged despite the satisfactory results obtained.

In the Gu et al. [13] study, a radiomics model was created for the noninvasive prediction of cytokeratin-19, galectin-3, and thyroperoxidase based on CT images. The study focused on the prediction of immunohistochemistry (IHC) of suspected thyroid nodules using ML-based radiomics. 103 patients with thyroid nodules were analyzed for the study. The cytokeratin 19 model achieved accuracy of 80.0% in the validation cohort. The thyroperoxidase and galectin 3 models attained accuracies of 84.2% and 85.0% in validation. In contrast, the high-molecular-weight cytokeratin model underperformed at 65.7% accuracy and lacked validation Moreover, it has been determined that thyroperoxidase immunoexpression positivity is a separate risk factor for thyroid cancer.

Zhou et al. [17] studied the radiomics analysis of iodine maps from dual-energy CT (DECT) to predict CLNM in PTC. In order to predict CLNM, two radiomics nomograms were created and validated. Nomogram 1 included LN rad-score, age (≥55), and CT-reported LN status, while Nomogram 2 incorporated CLN rad-score, capsule contact >25%, and CT-reported CLN status. Both nomograms demonstrated strong performance with AUCs of 0.847 and 0.837 in the training cohort and 0.807 and 0.795 in the validation cohort. Significant improvements in AUC and other metrics confirmed the added predictive value of the rad-scores. Both nomograms also showed good efficacy in CT-reported LN or CLN negative subgroups, with AUCs of 0.766 and 0.744. These nomograms demonstrated the promise of radiomics-based risk stratification in PTC, especially in CT-reported negative subgroups where they demonstrated favorable predictive efficacy. The study compared the predictive efficacy of radiomics nomograms and clinical models using DECT scanning with post-processing for radiomics analysis.

Lu et al. [12] presented a radiomics-based approach to predict cervical lymph node metastases in PTC patients prior to surgery. The study (involving 221 PTC patients) selects a radiomic signature with improved predictive performance by extracting 546 radiomic features from CT images. The radiomic signature outperformed individual radiomic sub-signatures created by SVM. Adding the radiomic signature to the nomogram enhanced predictive value, increasing the AUC from 0.807 to 0.867 in the training cohort, and confirmed in the validation cohort with an increase from 0.795 to 0.822.A prospective study by Koster et al. [15] sought to improve radiomics and [18F] FDG-PET/CT assessment in the differentiation of indeterminate thyroid nodules. AUC values for various SUV metrics range from 0.705 to 0.757. According to radiomic analysis, the PET model's AUC was 0.445. Although the study recommends higher SUV cut-offs for the best rule-out performance in Hürthle cell nodules, specificity data for quantitative [18F] FDG-PET/CT evaluation is absent In another study by Zhou et al. [14], of 255 lymph nodes (143 non-metastatic, 112 metastatic), radiomics features were extracted from iodine maps, and a radiomics signature was developed using LASSO logistic regression. Three models were created: model 1 (CT features), model 2 (radiomics signature), and model 3 (combined). The combined model (AUC = 0.933 training; AUC = 0.895 validation) outperformed both model 2 (AUC = 0.910 training; AUC = 0.847 validation) and model 1 in both sets. Z. Li et al. [122], in a study examined the effectiveness of a ML-based model in distinguishing nodular goiter (NG) from PTC using 3D arterial-phase contrast-enhanced computed tomography (CECT) features. Analyzing data from 407 patients and extracting 1,218 features, the radiomics model achieved an AUC of 0.877 in the validation group, outperforming clinical diagnosis, which had an AUC of 0.889 in the training group. With a classification accuracy of 80.49%, the results suggest that integrating advanced radiomics significantly enhances preoperative diagnostic accuracy for thyroid nodules. H. Xu et al. [21] in a rerospective studey investigated 280 patients with PTC. Recurrence was defined as confirmed disease or imaging evidence within 5 years post-surgery. Models were built for radiomics, clinical, and combined approaches using LR, SVM, KNN, and NN. Recurrence occurred in 89 patients, with 1,218 radiomics features extracted. Five radiomics and six clinical features were linked to recurrence. The LR and SVM radiomics models outperformed the NN model. Among clinical models, only the AUC difference between LR and NN was significant. Combined models showed higher AUCs than individual models, with validation cohort AUCs of 0.746 (LR), 0.754 (SVM), 0.669 (KNN), and 0.711 (NN). Also, Zhang et al. [23] collected CT imaging and pathology data from 141 patients with TN. From these, 1,316 radiomics features were extracted and refined using correlation testing, LASSO, or minimum redundancy maximum relevance selection. Four machine learning models (RF Hybrid Feature, RF, SVM Hybrid Feature, and LASSO) were developed and evaluated with ROC analysis. The RF Hybrid Feature model performed best, achieving an AUC of 0.87, significantly improving over the LASSO model (AUC = 0.78) in reclassification and discrimination metrics.

In a retrospective study by J. Huang et al. [22] a total of 376 cases collected and were divided into training and testing groups. Two radiologists identified ROI in the images and extracted 1,316 radiomic features. After feature screening, 11 optimal radiomic features were selected using mRMR and LASSO to develop an automated tumor discrimination model. The model's performance, assessed via ROC curve analysis, achieved an AUC of 0.911 on the training set, with accuracy, sensitivity, and specificity of 0.897, 0.889, and 0.862, respectively.

In summary, the highest performing radiomics model in CT images based on AUC was the combined radiomics model developed by Zhou et al. [14], with AUCs of 0.933 in the training set and 0.895 in the validation set, showcasing the effectiveness of integrating multiple features and methodologies in enhancing predictive accuracy for lymph node metastasis in PTC. Overall, these studies collectively underline the growing recognition of radiomics as a powerful tool in personalized medicine for PTC, each contributing to advancing the understanding of tumor behavior and improving preoperative assessments.

**6.3. Radiomics and AI using US Data for Thyroid Cancer** The remaining studies in the review exclusively utilized US images as the primary imaging modality for radiomics analysis in thyroid cancer diagnosis and prognosis. US imaging is widely utilized in clinical practice due to its non-invasive nature, accessibility, and ability to provide real-time imaging with high resolution [119]. Leveraging US images for radiomics analysis offers valuable insights into the morphological and textural features of thyroid nodules, facilitating the development of predictive models for risk stratification and treatment decision-making.

DL Radiomics of Thyroid (DLRT) was investigated by Zhou et al. [25] as a novel method for using US images to diagnose thyroid nodules. DLRT achieved the highest AUC and sensitivity across multiple validation cohorts, outperforming other DL models and even human observers in terms of accuracy. AUCs of DLRT were 0.96, 0.95, and 0.97 in the training, internal, and external validation cohort, respectively. DLRT performed significantly better than other DL models and human observers, Also, DLRT yielded the highest diagnostic accuracy for benign and malignant thyroid nodules. Sensitivities of DLRT reached 90.1, 89.3, and 89.5 in different cohorts. DLRT had higher accuracy compared to TL model and basic CNN model. This novel approach does away with the requirement for manual segmentation, makes use of TL to increase accuracy, and works well with a variety of US systems. The specificity of the model is not stated in the study, and more research is necessary to rule out any biases brought on by specimen collection and operator experience. Using US in radiomics to predict BRAF mutation in papillary thyroid carcinoma was investigated by Kwon et al, [29] who found that the technique was at most moderately successful. Even with an AUC of 0.65 and an average accuracy of 64.3%, it was not a very good diagnostic tool. Several ML models were used in the study, but none proved to be particularly beneficial. The authors referred to several drawbacks, including a small sample size, the possibility of overfitting the models, and intrinsic difficulties in quantifying US data in comparison to other imaging modalities. They draw the conclusion that the current utility of radiomics using US to predict BRAF mutation is limited, and they advocate for more study with larger datasets and a wider range of diagnostic techniques to validate and improve the methodology.

Göreke et al.'s [39] study developed a unique DL-based Computer-Aided Diagnosis (CADx) architecture for the use of US images in the classification of thyroid nodules. Using a multi-layer deep neural network (DNN) that combines morphological and texture features with features taken from US images via a pre-trained DNN, the suggested system outperforms existing literature in terms of performance. Recurrent Neural Network (RNN) architecture was chosen for the DL classifier due to its highest accuracy, and other noteworthy innovations include the addition of a pre-weighting layer using a modified genetic algorithm and a novel feature extraction method based on class similarity. Utilizing B-mode thyroid US images that have been manually segmented by a radiologist, the study includes a thorough comparison with previous research, bolstered by graphs and tables that demonstrate the validity and reliability of the suggested model. The efficacy of the suggested multi-layered classifier architecture is indicated by the AUC metric, which is the basis for the performance evaluation. Using a DNN that has already been trained on the ImageNet dataset, TL is used. The study's sample size is 1254 thyroid sonographic images. It focuses on the diagnosis of thyroid cancer and uses a variety of cutting-edge techniques, including deep active contour modeling for nodule segmentation, SNDRLS for nodule identification, and a Transformer and CNN-based strategy for the segmentation of malignant thyroid lesions within a CAD system.

Additionally, Zhu et al. [36] describe an US radiomics approach based on ANNs that predicts large-volume LNM in patients with clinical N0 PTC. The study uses deidentified data and a retrospective methodology, utilizing ANN-based the integrated model further enhances performance with an AUROC of 0.910, while the ANN-based radiomic model achieves an AUROC of 0.856. With an accuracy of 91%, the integrated model outperforms the radiomic model, which has an accuracy of 86%. Nevertheless, the information that is currently available lacks specifics for both the integrated model and the ANN-based radiomic model.

Xia et al. [30] report a study utilizing an AI algorithm to predict LNM in patients with PTC prior to surgery. The AI algorithm uses clinico-pathologic data to predict LNM by combining probabilistic neural network (PNN) and SVM. With an F1 score of 0.88 for central LNM and 0.93 for lateral LNM, the study illustrates the utility of this combination. Before surgery, the combination of AI algorithms and clinico-pathologic data is a useful tool for predicting LNM in PTC patients. AI algorithms in particular, which are predictive models, are useful for risk stratification.

Agyekum et al. [31] used a ML-based Clinical-Ultrasound Radiomic (Clinical-USR) model to assess CLNM in cases of PTC. The clinical-USR model outperforms an experienced sonographer in terms of accuracy when predicting CLNM in PTC patients. The model considers PTC heterogeneity, a measurable characteristic linked to malignancy, which eliminates subjectivity in standard US imaging diagnosis. 205 patients were included in the retrospective study. Nine statistically significant variables were used to build the Clinical-USR model, with internal echo, shape, age, and mass being identified as the main predictors. The Clinical-USR model distinguished PTC patients with CLNM, showing AUCs of 0.78 (training) and 0.71 (validation). It outperformed the Clinical model, particularly in validation (0.71 vs. 0.63). The research notes certain difficulties, including the challenge of using ultrasonography to detect all metastatic lymph nodes, subjectivity in standard ultrasonography diagnosis, and limited visual measurement of PTC density and tissue heterogeneity.

In Wang et al.'s [26] study, thyroid nodule classification using US images was compared between radiomics and DL techniques. The study used a dataset consisting of 3120 US images from 1040 cases. Interestingly, DL performed better than radiomics, with an accuracy of 74.69% against radiomics' 66.81%. Large datasets are essential for improving the performance of DL models, according to the research. It was suggested that future work focus on DL model optimization and multicenter data validation of efficacy..

The goal of Yu et al.'s [33] research is to predict RET rearrangement in PTC by means of a novel method that combines DL and US radiomics. AUCs of 0.9545 in the test cohort and 0.9396 in the training cohort indicate the excellent performance of the DL radiomics nomogram (DLRN), which was developed. The behavior of aggressive PTCs, RET rearrangement, was predicted by using ultrasonic radiomics and deep transfer learning. Among the clinical characteristics that separated patients with and without RET rearrangement were age, tumor size, and calcification.

Wang et al.. [32] was attempted to predict BRAF V600E mutations in PTC using grayscale imaging and elasticity US. The development of radiomic models from US images showed high accuracy in mutation prediction, with areas AUC ranging from 0.725 to 0.985. The model with the highest AUC was the combination of elasticity and grayscale. 138 PTC patients underwent preoperative ultrasonography, and radiomic features were taken out for model creation. Some of the limitations were the small sample size, the retrospective design, and the absence of outside validation. .

By utilizing an US radiomics nomogram, the Zhou et al. [34] study presents a novel method for preoperative prediction of central neck LN metastasis in papillary PTC. The study integrates radiomics features, US findings, and clinical factors to develop a predictive model in response to the suboptimal accuracy of US in diagnosing central cervical LN metastasis in PTC. In the validation cohort, the area under the AUC was 0.858. With its high AUC values and remarkable sensitivity, specificity, and accuracy.

An US-based radiomics XGBoost model is created in the Shi et al. [35] study to predict central cervical lymph node metastasis (CCLNM) in PTC patients. The model uses 11 radiomics features, including important variables like capsular invasion, radiomics score, diameter, age, and calcification, to calculate the radiomics score after retrospectively analyzing data from 587 PTC patients. When compared to radiologists, the XGBoost model performs better, showing an astounding 44% increase in AUC. The AUC was 91.53% and 90.88% in the training and test sets, respectively. The stability of the XGBoost model is particularly noteworthy, as it demonstrates reliable performance on various kinds of US scanners. The study highlights the importance of CCLNM preoperative prediction in PTC and shows that US-based radiomics, XGBoost, and SHAP are effective tools for precise risk assessment and customized interpretation of predictive features in the diagnosis of thyroid cancer. The XGBoost model's positive and negative effects on each parameter are revealed by the SHAP plots, which enhances the model's interpretability and comprehension of its decision-making process. The extensive methodology emphasizes the integration of cutting-edge techniques in thyroid cancer research and includes image segmentation, feature extraction, normalization, LASSO logistic regression, and the creation of the XGBoost model.

In a study Li and colleagues [37] used a multi classifier US radiomics model to develop a ML technique for predicting ETE in pediatric PTC. Four supervised ML models were created: LightGBM, k-nn, RF, SVM, and LightGBM. The LightGBM model performed exceptionally well in both the training and validation cohorts. In the training cohort, AUCs were 0.880, 0.873, 0.999, and 0.926 for SVM, KNN, RF, and LightGBM, respectively. In the validation cohort, AUCs were 0.784 (SVM), 0.720 (KNN), 0.728 (RF), and 0.832 (LightGBM). The amalgamated model, which fused ultrasonic radiomics and ML, demonstrated exceptional prognostic potential for ETE in pediatric PTC. Using a correlation coefficient screening method, ML models were developed, radiomics features were extracted from US images, and the SHAPley Additive explanations SHAP framework was used to explain the models.

In a study, Kwon et al. [27] used radiomics analysis based on gray-scale US to predict distant metastasis in FTC. The second most common thyroid cancer is called FTC, and a major predictor of a poor prognosis is distant metastasis. The objective of the study was to create an independent biomarker for distant metastasis in FTC using radiomics signatures. Radiomics analysis used the least absolute shrinkage and selection operator to create a radiomics signature by extracting features from US images. The radiomics signature was utilized to train an SVM classifier. The classifier showed an AUC of 0.90 on the test folds. With an AUC of 0.93, multivariate analysis showed the classifier's strong diagnostic performance Distant metastasis was linked to clinical factors like size, age, widely invasive histology, lymph node metastasis, and extrathyroidal extension. On the other hand, widely invasive histology and the radiomics signature were found to be independent variables linked to distant metastasis. Using US and histological images, Rohit Sharma et al. [38] provides a thorough framework for the detection of thyroid cancer that integrates DL, meta-heuristics, and Multiple Criteria Decision Making (MCDM) algorithms. The study evaluates five classifiers using three DL techniques and six feature transformation methods, and uses MEREC-TOPSIS MCDM for model ranking.. The PCA+FOX feature selection with RF achieved top performance, with a 99.13% accuracy, and 99.13% AUC-ROC on US data, and 90.65% accuracy, and 95.48% AUC-ROC on histopathology data. By underlining early diagnosis, the framework seeks to reduce the workload for medical practitioners by offering enhanced diagnostic capabilities. The study pointed the significance of accurate CAD models for patient care by utilizing pre-trained DL models, dimensionality reduction techniques, and the FOX optimization algorithm for feature selection The use of multiple imaging modalities, such as US, and the investigation of DL models, such as DeiT, Swin Transformer, and Mixer-MLP, for feature extraction are important components of their methodology. In addition, the novel FOX optimization algorithm for feature selection and ensemble learning was used, along with dimensionality reduction techniques such as PCA and ISOMAP. Though restricted by the availability of publicly available standard thyroid datasets, the framework shows promising results despite concentrating on two particular thyroid image datasets.

Liu et al. [24], conducted a study focusing on the prediction of LNM in patients with PTC using a radiomics approach based on preoperative US images. After 450 US images were examined, 50 features were found to have better accuracy in predicting the status of lymph nodes. With a 0.712 accuracy and a 0.782 AUROC, the proposed model appears to be a promising noninvasive method for assessing metastasis risk The method used a combined feature selection strategy, an SVM classifier for model building and validation, and 614 high-throughput features extracted from US images. Manual segmentation of tumors and a three-step feature selection method were employed to enhance the accuracy and reliability of the model. The study's findings contribute to the growing body of evidence supporting the utility of radiomics in noninvasive cancer prediction and management, particularly in high-incidence cancers like PTC.

In a retrospective study, Zhu et al. [40] used ML techniques to create preoperative prediction models for evaluating LNM in thyroid cancer cases. The models combined clinical traits and ultrasonic radiomic features to help with diagnosis, predict recurrence, and guide treatment choices. With an accuracy of 0.7542 and an AUC of 0.812, the RF-RF model demonstrated the best predictive performance for cervical

LNM among the five classifiers that were used. Notably, the most important factors influencing the effectiveness of the RF-RF model were found to be age and a variety of radiomic features. The superior performance of the RF-RF model was demonstrated by using Delong's test to compare its AUC with other models.

Yu et al. [28], presents a transfer learning radiomics (TLR) model for predicting LNM in PTC patients, achieving a stable LNM prediction with an average AUC of 0.90 across three datasets totaling 3,172 patients. Decision curve analysis shows that the TLR model performs better than other approaches, including DL models such as VGG, ResNet, and Inception ResNet, and benefits PTC patients more. It can be used to guide treatment decisions for PTC patients and improve diagnostic performance. X. Keutgen et al. [6] labeled Ultrasound images from two institutions by FNA and surgical pathology as malignant (M), benign (B), or indeterminate (I), resulting in subgroups: 90 BB, 83 IB, 70 MM, and 59 IM thyroid nodules. Thyroid nodule margins were manually annotated, and radiomic texture analysis was performed within these contours. Using five-fold cross-validation and a two-class Bayesian neural network classifier with stepwise feature selection, the model was validated on a separate test set and compared to a commercial molecular testing platform. AUCs were 0.75 for MM vs. BB and 0.67 for IM vs. IB in the training/validation set of 263 nodules. For an independent test set of 19 IM/IB nodules, the algorithm achieved an AUC of 0.88, surpassing the commercial platform's AUC of 0.81. A recent retrospective study by Y. Gao et al. [45] analyzed 613 PTC patients, developing a DL and radiomics model using primary lesion images, and integrating clinical and US features through multivariate analysis. Compared to junior and senior radiologists, the integrated model achieved the highest AUC (0.841), surpassing the radiomics (0.706) and DL models (0.819). It lowered the missed CLNM rate to 27.63% and reduced unnecessary CLND to 18.18%. SHAP analysis showed DL features were central in diagnosing CLNM, with clinical and US features enhancing accuracy.

In a study by F. Fan et al. [46], 211 lymph nodes from postoperative differentiated thyroid carcinoma (DTC) patients, a predictive model for CLNM was developed. Using data split into training (n=147) and test (n=64) cohorts, relevant radiomic features from US images were identified with the LASSO algorithm to build a radiomics model. The combined model—incorporating radiomics, ultrasound, and clinicopathological data—achieved an AUC of 0.901 in predicting CLNM in the test cohort, significantly outperforming the single ultrasound model (AUC 0.731) and radiomics model alone (AUC 0.801). Z. Ni et al. [47] In a multicenter retrospective study of 854 PTC patients, radiomics and DL models were built using multiple algorithms, including ExtraTrees, DenseNet121, and ResNet50. The ExtraTrees model achieved the highest AUC of 0.787 among radiomics models, while the DenseNet121 DL model reached 0.766. The combined Thy-DL-Radiomics model, integrating radiomics and DL features, showed improved diagnostic performance with an AUC of 0.839 in internal validation and 0.789 in external validation. Also, M. Arabi et al. [42] investigated 210 patients with thyroid nodules, analyzed radiomics features from US images using LASSO, MRMR, and XGBoost for feature selection and classification. GLRLM-RLNU and GLZLM-GLNU achieved the top AUCs (0.67) in univariate analysis for malignancy prediction. In multivariate analysis, the XGBoost classifier combined with MRMR showed the highest performance, achieving an AUC of 0.99 in training and 0.95 in testing. Similarly, F. Ruqian et al. [48] examined 528 patients with Hashimoto's thyroiditis (HT) and PTC, groups were formed based on the presence LNM. Three US specialists extracted radiomic features, which were combined with clinical data to build five machine learning models: random forest, SVM, LightGBM, KNN, and XGBoost. Model performance was assessed with ROC curves, with LightGBM and XGBoost achieving the highest AUC scores (0.798–0.921) and outperforming others ($P<0.05$). Models combining radiomic and clinical features showed improved accuracy ($P<0.05$). SHAP analysis pointed features like anteroposterior and superoinferior diameters, voxel volume, and age as key predictors for both top models. H. Du et al. [49] included 1,076 thyroid nodules from 817 patients across three institutions. Radiomics and DL features were extracted from US images to create radiomics (Rad_sig) and deep-learning signatures (DL_sig). A clinical ultrasound semantic signature (C_US_sig) was also developed. A combined nomogram model was built and evaluated. The model achieved AUCs of 0.947, 0.917, and 0.929, demonstrating superior clinical utility in decision curve analysis and good calibration. AUCs for distinguishing benign and malignant nodules ranged from 0.714–0.752 for junior physicians to 0.891–0.908 for experts, improving to 0.858–0.923, 0.888–0.944, and 0.912–0.919 with DL radiomics assistance.

A study by T. Zhou et al. [43] analyzed 192 patients with 197 partially cystic thyroid nodules (PCTNs). Radiomics features were extracted from US images, and ML methods were used to build a classification model. Model performance was evaluated using AUC, accuracy, sensitivity, and specificity. Nineteen radiomics features were extracted for each nodule, and eight ML classifiers differentiated malignancy in PCTNs. The k-NN model achieved an AUC of 0.909, with accuracy, sensitivity, and specificity of 82.95%, 83.33%, and 89.90%, respectively. The ML-based US radiomics system outperformed the ACR TI-RADS model and US features model in AUC. H. Zhao et al. [50] investigated the patients scheduled for surgical treatment due to thyroid cancer were prospectively included. Each patient underwent US of CLNs suspected of metastasis, with a 2-minute imaging video recorded. After target tracking and feature extraction, three ML models—SVM, linear discriminant analysis (LDA), and DT—were developed. A total of 75 lymph nodes were analyzed, with SVM showing the best diagnostic efficacy: 93.0% sensitivity, 93.8% specificity, and 93.3% accuracy, calculated via leave-one-out cross-validation (LOOCV). In another study designed by Q. Dai et al. [41], 498 unifocal PTC cases with post-operative histopathology as the CLNM standard. US images were segmented, and multimodal features were extracted using a 50-layer DNN. Features were then selected and fused, followed by classification using adaptive boosting (AB), LDA, and SVM. Among the models, Multimodal-SVM performed best ($P < 0.05$), achieving AUCs of 0.910 (validation) and 0.851 (testing), with subgroup AUCs of 0.920 in cN0 subgroup-1 and 0.828 in cN0 subgroup-2.

In this review, most studies utilized US images as the primary modality for radiomics analysis in thyroid cancers, reflecting its widespread availability and effectiveness in clinical practice. The procedures include a wide range of DL techniques, like CNNs and TL, which have shown remarkable promise in enhancing diagnostic accuracy. The TL model performed impressively across multiple cohorts, pointing its strength and potential for broad clinical application.

Traditional ML models, such as SVM and RF, showed more variable results, reflecting a variety of performance levels. This variation means that, while certain models may perform well in specific scenarios, they will be struggling to overcome the complexity of US data and the heterogeneity of thyroid cancers. likewise models that included clinical and radiomic features with imaging data provided better outcomes, showing the need of a multimodal approach. This integration can combine quantitative data from radiomics with qualitative insights from clinical assessments to create a broader view of patient risk. However, some studies discovered lower accuracy, outlining the persistent problems in developing and validating radiomics models across patient groups and imaging conditions.  These discrepancies raise important questions about the reproducibility and generalizability of current models in clinical settings. The findings underscore the promising role of AI and radiomics in improving diagnostic accuracy for thyroid cancer, yet they also point to the necessity for standardized methodologies and larger, more diverse datasets to enhance the reliability and applicability of these models in everyday clinical practice. Reviewing radiomics in thyroid cancers, especially when combined with AI applications, reveals a wide range of results and approaches. On the other hand, a radiomics study that used US to predict BRAF mutation in thyroid nodules performed moderately well, with average accuracy, sensitivity, and specificity of 64.3%, 66.8%, On the other hand, a radiomics study with an average accuracy, sensitivity, and specificity of 64.3%, 66.8%, and 61.8%, respectively, demonstrated a moderate level of success in using US to predict BRAF mutation in thyroid nodules. and 61.8%, respectively. Notably, different feature-selection strategies did not significantly improve test performance in this particular study. Furthermore, different classifiers performed differently in different studies, highlighting the variation in results [29].

In contrast, some studies introduced novel approaches that demonstrated superior performance compared to existing literature. For instance, the application of a multi-layered classifier architecture, particularly utilizing RNN, exhibited high accuracy in comparison to traditional models [39]. Comparably, enhanced discrimination and clinical utility were demonstrated by the incorporation of artificial neural network-based US radiomics for forecasting large-volume lymph node metastasis in patients with papillary thyroid carcinoma [36, 52]. These results highlight how customized AI applications can improve the precision of diagnoses in particular clinical situations.

The high F1 scores that the PNN and SVM classifiers achieved in different test sets [30] showed how well AI algorithms combined with clinico-pathologic data could predict lymph node metastasis in cases of papillary thyroid carcinoma. This was the subject of another study. To fully realize the potential of the technology, a collaborative approach was suggested by emphasizing the need for a supervisory framework for AI applications that includes radiologists [123]. Moreover, given the variability of papillary thyroid carcinoma, the clinical-USR model outperforms a skilled sonographer in assessing central lymph node metastasis [31], highlighting the developing role of AI in improving diagnostic accuracy. When comparing DL with radiomics-based methods, different results were also observed. DL consistently outperformed radiomics-based methods in terms of accuracy and AUC in predicting different parameters, including BRAF V600E mutations [26] and RET rearrangement [33]. These findings imply that, for some diagnostic tasks, DL techniques may become more advantageous than traditional radiomics. The literature also pointed how radiomics is a flexible tool for predicting different aspects of thyroid cancers, including metastasis to central neck lymph nodes [35], distant metastasis in FTC [27], and ETE [19]. This versatile application demonstrates how radiomics can offer comprehensive insights into the features and prognosis of thyroid cancers.

In conclusion, there is a wide range of performance and utility levels in the literature on radiomics in thyroid cancers and AI applications. While some studies show encouraging outcomes and improvements in diagnostic accuracy, others draw attention to the necessity of carefully weighing the unique clinical context and cooperatively integrating AI into already-existing diagnostic frameworks [124]. The inconsistent results point to the significance of individualized strategies and ongoing improvement in the use of AI and radiomics for thyroid cancer prognosis and diagnosis.

**7. Staging**

Based on publication of the eighth edition of the American Joint Committee on Cancer (AJCC)/Union for International Cancer Control (UICC) in 2016, changes were made to the staging system for thyroid cancer [125]. Remarkably, age was taken into consideration; the age cutoff was 55 instead of 45. Significant downstaging of tumor classifications resulted from the revisions; 38% of patients had their tumor classifications downstaged, and the percentage of patients classified as Stage I or II increased from 64% to 94%. These modifications provided a more accurate prognosis by better aligning staging with disease-specific survival. The American Thyroid Association's (ATA) risk stratification system and AJCC staging together enhanced the prediction of disease-specific survival [126]. Detectable thyroid nodules are present in only 30–40% of thyroid cancer instances, rendering them a secondary diagnostic preference. Essential for identifying thyroid nodules are imaging techniques like cross-sectional and ultrasound imaging [127]. However, the concern of overdiagnosis underscores the importance of exercising prudence and adhering to evidence-based guidelines discouraging screening for asymptomatic individuals. Ultrasound emerges as the favored imaging modality for evaluating nodule characteristics, though its application requires circumspection to prevent misdiagnoses. Presently, fine needle aspiration (FNA) remains a dependable diagnostic approach, particularly when coupled with risk assessment tools [128]. Molecular diagnostic methodologies, exemplified by ThyroSeq v3 and Afirma, have mitigated the necessity for diagnostic lobectomy in cases of indeterminate nodules. Striking a judicious balance between early detection and the risk of overdiagnosis persists as a formidable challenge confronting medical practitioners [129].

## 8. Challenges, Limitations and Future Directions

Obstacles and restrictions in the use of radiomics and AI in the detection and prognosis of thyroid cancer. A persistent issue pertains to the dependence on operator expertise and specimen gathering, which may introduce errors in production, underscoring the necessity of standardization and careful handling [25]. There are inherent limitations with US-based radiomics, especially in predicting the status of BRAF mutations. These include operator dependence, difficulties with quantitative analysis, and the need for larger datasets to improve the efficacy of particular features [29]. In order to achieve more reliable and broadly applicable results, prospective multicenter approaches to validation are crucial, as evidenced by the retrospective nature of certain studies and the uneven distribution of data [36]. One major issue that arises is the interpretability of AI models, since classifiers are "black boxes," making it difficult to understand the causal relationships between them. Variations in tumor microenvironments and imaging protocols make generalizing models even more difficult, highlighting the necessity of fully comprehending the underlying concepts and incorporating clinical data [34, 110]. Even with their promising performance, AI systems perform on par with or worse than experienced ultrasonography specialists, highlighting the necessity of a supervisory structure for radiologists as of right now [123]. Problems with the dataset include small sample sizes, single-center designs, and low diversity, which emphasizes the significance of using multi-modal, multi-center data for thorough validation [18, 27, 32, 33]. Different imaging modalities, like [18F]FDG-PET/CT, present difficulties for radiomics methodology due to limited specificity and difficult differentiation [15]. Similar challenges have been reported in medical image analysis, including the need for large annotated datasets, reduced performance in sequential data, computational inefficiency, and limited generalizability across imaging modalities[130].

As seen by the difficulties in US radiomics, a common theme in radiomics models is a lack of standardization. It will take work to standardize the field, cut down on mistakes, and improve US radiomics' applicability in a variety of clinical contexts [27]. The necessity for additional validation in prospective datasets, with careful consideration of potential biases, is highlighted by retrospective study designs [12]. The difficulties in making clinical decisions like the erratic US physicians' rating skills and the lack of certainty in cytopathological diagnosis—highlight how difficult it is to incorporate these technologies into real-world medical settings. Additionally, limitations in the knowledge base for head and neck diseases, coupled with a scarcity of studies on radiomics in certain thyroid cancers, the need for comprehensive research in these areas[88].To overcome the challenges and limitations in the application of radiomics and AI for thyroid cancer detection and prognosis, concerted efforts and innovative solutions will be required. Standardization efforts are essential for enhancing reliability and repeatability in a range of clinical settings, and multimodal data integration presents the possibility of developing comprehensive predictive models. Advancements in explainable AI (XAI) technology have the potential to improve the interpretability of ML models, thereby simplifying their integration into clinical environments. Prospective multicenter studies with larger datasets are necessary to validate the efficacy and generalizability of radiomics models, and personalized medicine approaches can customize treatment regimens to the specific features of individual tumors. The creation of clinical decision support systems and funding for training and education programs for medical professionals are crucial for the widespread adoption and application of radiomics-based approaches. By embracing these new directions, the field will be able to advance and enhance patient outcomes in the treatment of thyroid cancer by facilitating more accurate diagnosis and customized treatment.

## 9. Conclusion

This review presents a landscape of changing approaches and results from a thorough analysis of the literature on the use of AI and radiomics in thyroid cancer diagnosis and prognosis. Due to its widespread availability and efficacy in clinical practice, US images have been the primary modality for radiomics analysis in the majority of studies in this domain. Notably, combining computational power with state-of-the-art imaging techniques has shown promise in improving accuracy, effectiveness, and customized patient care. Numerous approaches and methodologies have been used in the reviewed studies, producing a wide range of outcomes and results. While some research has presented innovative techniques that outperform conventional approaches, other studies have emphasized the difficulties and restrictions that come with using AI and radiomics to diagnose thyroid cancer. Research employing multi-layered classifier architectures, specifically with RNN, has demonstrated superior accuracy in certain clinical situations. Furthermore, integrating artificial neural network-based US radiomics has demonstrated potential in predicting parameters like large-volume lymph node metastasis in papillary thyroid cancer patients. But problems like operator dependence, AI model interpretability problems, dataset limitations, and the ongoing need for standardization efforts still exist. Furthermore, variations in performance and outcomes amongst studies highlight the necessity of additional methodology validation and improvement. Different results have been observed when comparing radiomics-based methods, with DL consistently outperforming radiomics-based methods in some diagnostic tasks. This implies that in some situations, DL techniques might be more advantageous than traditional radiomics. Despite the variability in results and performance levels observed in the literature, radiomics appears to be a flexible tool capable of predicting various aspects of thyroid cancers, including metastasis to central neck lymph nodes, distant metastasis in follicular thyroid cancer, and extrathyroidal extension. This adaptability emphasizes radiomics' potential to provide thorough insights into the characteristics and prognosis of thyroid cancers. In conclusion, there is considerable potential for these technologies to transform thyroid cancer diagnosis and prognosis, even though the literature on radiomics in thyroid cancers and AI applications shows a wide range of performance and utility levels. However, coordinated efforts in standardization, validation, and integration into clinical practice will be necessary to realize this potential. By taking on obstacles and utilizing creative solutions, the field can progress toward better patient outcomes and a more thorough comprehension of the pathophysiology of thyroid cancer.

## Declaration

### Ethics approval and consent to participate:

This study was conducted under the principles outlined in the Declaration of Helsinki.

### Consent for publication

N/A

### Availability of data

N/A

### Competing interests

The authors declare that there is no conflict of interest.

### Funding Sources

None.

**CRedit Statement**

Conceptualization, M.Y. and H.V.; methodology, S.F.M,; validation, M.A.Y, H.V.,S.P., and P.R.; formal analysis, M.Y. and H.V.; investigation, S.F.; resources, K.I.A.L.A.A.; writing—original draft preparation, H.V., M.Y., M.A.Y, A.J., P.R., and S.F.M.; writing—review and editing, R.T., P.P., K.I.A.L.A.A., S.P. and R.A.; visualization, M.Y.; supervision, K.I.A.L.A.A. and S.P.; project administration, H.V.; All authors have read and agreed to the published version of the manuscript.. Every author adhered to the COPE guidelines as universally acknowledged.

**Acknowledgments**

None.

| Table 1: overview of studies in field of integration of radiomics and artificial intelligence in thyroid cancers | | | | | |
|---|---|---|---|---|---|
| Reference | Imaging | Sample size | Tasks | Approaches | Performance |
| Hui Qin et al., 2021 [9] | MRI | MRI 109 patients with PTC | Classification (lymph node metastasis or not) | MRI-Texture (naked eye), MRI-Radiomics (LR model), Combination model, | MRI texture alone showed AUC 0.739, radiomics model AUCs 0.805 (training) and 0.760 (validation), and combined model AUC 0.969, indicating high diagnostic efficiency for LN metastasis in PTC. |
| Wei et al., 2021 [10] | MRI | MRI 132 patients with PTC | Binary Classification (those with ETE and those without ETE) | DWI, T2 WI, CE-T1WI | AUC 0.87 |
| Dai et al., 2022 [11] | MRI | MRI 107 patients with PTC | Classification (aggressiveness and non-aggressiveness) | meta-analysis- sparse representation method | CE-T1WI, T2WI and DWI combined 0.917 |
| Lu et al., 2019 [12] | CT | CT images from 221 PTC patients | Classification (central cervical lymph node metastasis) | SVM, Multivariate logistic regression | AUC of 0.795 and 0.822 respectively for SVM and multivariate logistic regression |
| Gu et al., 2019 [13] | CT | CT images 103 patients with suspected thyroid nodules | Classification (malignant or benign) | SVM | AUC 0.87, sensitivity 0.93, and specificity 0.73 |
| Zhou et al., 2020, [14] | CT | CT images from 255 PTC patients | Classification (Metastasis to cervical lymph node or not) | LR, Radiomics Signature Model, Combined Model | Combined model showed optimal performance in both training (AUC = 0.933) and validation set (AUC = 0.895) |
| de Koster et al., 2021 [15] | FDG-PET/CT | [18F] FDG-PET/CT thyroid nodule from 123 patients | Classification | SUV- Elastic net regression classifiers | SUV-metrices showed AUCs ranging from 0.705 to 0.729 in all, 0.708 to 0.757 in non-Hurthle and 0.533 to 0.700 in Hurthle cell |
| Li et al.,2021 [16] | CT | CT images of Thyroid 678 patients with PTC | Binary Classification ( CLNM-, CLNM+) | KNN, LR, DT, L-SVM, Polynomial-SVM, Gaussian SVM algorithm | L-SVM: higher classification accuracy 0.709 Internal and external validation c-index 0.854 |
| Zhou et al., 2022 [17] | CT | CT images from 346 patients with PTC | Classification (LNM, CLNM) | MRMR, RF, LASSO, LR, Decision Curve Analysis | discrimination and calibration of training cohorts (AUC = 0.847; AUC = 0.837) validation cohorts (AUC = 0.807; AUC = 0.795 |
| Wu et al.,2022 [18] | CT | CT images from 412 patients with PTC | Classification (TCI+, TCI-) | KNN, LR, DT, L-SVM, Polynomial-SVM, SVM | internal test cohort L-SVM, AUC=0.820 [95%CI 0.758–0.888] external test cohort, the AUC was 0.776 (0.625–0.904) |
| Yu et al., 2022 [19] | CT | CT images from 153 patients with PTC | Classification (extrathyroidal extension (ETE)) | LASSO, KNN, logistic regression, DT, linear -SVM, gaussian-SVM, and polynomial-SVM | Best internal test cohort: L-SVM, AUC=0.820 best radiomics model (L-SVM, AUC = 0.733) clinical model (AUC = 0.709 external test cohort, the AUC was 0.776 |
| Xu et al., 2022 [20] | CT | CT radiomics from 161 patients | Binary Classification (Malignant or benign) | LR | AUC of 94.4%, an accuracy of 87.6% in the train, and 94.2%, and 86.8% respectively for the test group |
| Xu et al. 2022 [21] | CT | CT radiomics from 280 patients | Predicting postsurgical recurrence of PTC | LR-, NN-, SVM-, and KNN- based radiomics models | The combined models based on the SVM, LR, KNN, and NN classifiers had AUCs of 0.754, 0.746, 0.669, and 0.711, respectively |
| J. Huang et al., 2023 [22] | CT | CT from 376 patients | Classification (HT or thyroid cancer) | LASSO, MRMR, | AUC) of 0.911, high accuracy, sensitivity, and specificity, measured at 0.897, 0.889, and 0.862, respectively |
| Zhang et al. 2024 [23] | CT | CT radiomics and pathological data from 141 patients with TN | Classification (NG or malignant nodule) | Radiomics- based four different model: RF, RF Hybrid Feature, SVM Hybrid Feature, and LASSO | The RF Hybrid Feature model with an AUC of 0.87 for the validation cohort, outperforming the LASSO model, which had an AUC of 0.78 |
| Liu et al., 2019 [24] | US | US images from 450 PTC patients | Classification (presence of lymph node metastasis or not) | SVM | AUORC of 0.727, accuracy of 0.710 |
| Zhou et al., 2020 [25] | US | US images from 1734 patients | Binary Classification (Malignant or benign) | CNN, TL, DLRT | AUCs of DLRT were 0.96, 0.95, 0.97 respectively in the training, internal and external validation cohort |
| Wang et al., 2020 [26] | US | US images from 1040 patients | Binary Classification (Malignant or benign) | Radiomics model, DL | Accuracy of 66.81% in radiomics and 74.69% in DL |
| Kwon et al., 2020 [27] | US | US radiomics from 169 patients with FTC | Classification (distant metastasis or not) | SVM classifier Radiomics and histological features | AUC of 0.90 for SVM, and 0.93 for radiomics and widely extensive histology feature |
| Yu et al., 2020 [28] | US | US images of 3172 patients with PTC | Classification of patients into groups based on the risk of lymph node metastasis in PTC | Transfer Learning-based model (TLR) | AUC main cohort: 0.90 independent cohorts:0.93 |
| Kwon et al., 2020 [29] | US | Gray-scale US from 96 patients | Binary Classification (presence or absence of BRAF mutation in PTC) | LR, SVM, RF | AUC of 0.651, accuracy of 64.3%, sensitivity of 66.8%, and specificity of 61.8%, on average, for the 3 models |
| Xia et al., 2021 [30] | US | Clinical and US data from 445 PTC patients | Classification (presence of lymph node metastasis or not) | SVM, PNN | F1 score of 0.88 on the central lymph node metastasis in PNN, F1 score of 0.93 on the lateral lymph node metastasis in SVM |
| Agyekum et al., 2022 [31] | US | Clinical and US data from 205 PTC patients | Classification (cervical lymph node metastasis) | USR model<br>Clinical model,<br>USR-Clinical model | AUC of 0.71 in USR-clinical, 0.74 in USR, and 0.63 in clinical model |
| Wang et al.,2022 [32] | US | US images from 138 Patients with PTC | Binary Classification (BRAF+, BRAF-) | LASSO-mRMR- | Elasticity US AUC 0.952, grayscale US AUC 0.792, and combination AUC 0.914 in the training dataset, elasticity US AUC 0.931, grayscale US AUC 0. 725 and combination AUC 0.938 in the test dataset |
| Yu et al.,2022 [33] | US | US images from 650 patients with PTC | Binary Classification (RET/PTC, Non-RET/PTC) | DLRN<br>logistic regression algorithm | AUC of DLRN in test cohort 0.9545 and in training cohort 0.9396 |
| Zhou et al.,2022 [34] | US | US images of thyroid and central cervical lymph from 609 patients with PTC | Classification | Lasso regression model -multivariable logistic regression radiomics signature, ultrasonography-reported LN | C-index of 0.816 in the primary cohort and 0.858 in the validation cohort<br>AUC 0.858. In the validation cohort, the accuracy, sensitivity, specificity and AUC of this model were 0.812, 0.816, 0.810, and 0.858 |
| Shi et al., 2022 [35] | US | US imaging, clinical data, and radiomics of 587 PTC patients | Classification (central cervical lymph node metastasis) | Boruta algorithm,<br>XGBoost model, | AUC 91.53% |
| Zhu et al., 2022 [36] | US | US images from 306 PTC patients | Classification (Large volume lymph node metastasis (LNM)) | ANN based radiomic model | AUROC of 0.856 and AUPR of 0.381 for radiomic model, AUROC of 0.910 and an AUPR of 0.463 for integrated model |
| Keutgen et al.2022 [6] | US | 1052 ultrasound images from 302 thyroid nodules | Binary Classification (Malignant or benign) | two-class Bayesian ANN classifier | Achieved AUCs of 0.75 for MM vs. BB and 0.67 for IM vs. IB. On an independent test of 19 IM/IB cases, it reached an AUC of 0.88, outperforming a commercial molecular test (AUC = 0.81) |
| Li et al., 2023 [37] | US | US images from 164 pediatric patients with PTC | classification (extrathyroidal extension or not) | LightGBM, KNN, SVM, RF | AUC of 0.832, 0.720, 0.784, 0.728 respectively for four model |
| Sharma et al., 2023 [38] | US | Two datasets: the US dataset (298 patients) and the Histopathological Dataset (156 thyroid tumors) | Classification (malignant or benign) | transformation techniques (PCA, TSVD, FastICA, ISOMAP, LLE, and UMP)<br>and (LR, NB, SVC, KNN, RF) | RF and PCA+FOX yielded the best performance: AUC of 99.13%, an accuracy of 99.13% in the US, and 95.48%, and 90.65% respectively in the histopathological dataset |
| Göreke et al., 2023 [39] | US | B-mode thyroid US images from 99 patients | Binary Classification (Malignant or benign) | ANN, CNN, LSTM, RNN | AUC of 1 and recall of 0.99 for all of them, |
| Zhu et al., 2023 [40] | US | US images from 371 PTC patients | Binary Classification (with cervical lymph node metastasis, without cervical lymph node metastasis) | RF, LASSO, binary LR, KNN, SVM, XGboost | AUC of RF-RF model validation group (0.812) |
| Q. Dai et al., 2023 [41] | US | US data from 498 PTC patients | Binary Classification (CLNM or not) | AB, LDA, and SVM- based radiomics model | The Multimodal-SVM model showed AUCs of 0.910 in validation, 0.851 in testing, 0.920 in cN0 subgroup-1, and 0.828 (95% CI, 0.769–0.887) in cN0 subgroup-2 |
| M. Arabi et al., 2023 [42] | US | US data from 210 patients | Prediction (Malignant or benign) | Lasso, MRMR, RF, XGBoost | GLRLM-RLNU and GLZLM-GLNU showed top AUCs of 0.67 in predicting nodule malignancy. On the test data, XGBoost with MRMR and LASSO selection achieved the highest performance with an AUC of 0.95. |
| T. Zhou et al., 2023 [43] | US | US data from 192 patients | Classification (malignant or benign) | KNN- based radiomics model, other seven ML classifiers | The AUC, accuracy, sensitivity, and specificity of KNN model were 0.909, 82.95%, 83.33%, and 89.90%, respectively |
| J. Cozzi et al., 2023 [44] | US | 476 US images from patients with intermediate nodules | Classification of Intermediate Nodules (Final Dx: Benign or Malignant) | radiomics and DL based models: classifier combination model (SCM), interpretability-driven combination model (ICM) | The models achieved mean AUCs of 0.75 for RM, 0.70 for DTLM, 0.77 for SCM, and 0.76 for ICM. |
| Y.Gao et al., 2024 [45] | US | US and clinical data from 613 PTC patients | Classification (CLNM, non-CLNM) | hand-crafted radiomics model, DL-based Radiomics model, integrated model | The integrated model achieved the highest AUC (0.841), surpassing both the radiomics (0.706) and DL models (0.819). |
| F. Fan et al., 2023 [46] | US | US data from 211 DTC patients | Classification (CLNM or not) | ML-based Radiomics model | The combined model achieved an AUC of 0.901, significantly outperforming the single ultrasound model (AUC 0.731) and radiomics model alone (AUC 0.801). |
| Z. Ni et al., 2024 [47] | US | US data from 854 PTC patients from three different center | Classification (large volume lymph node metastasis or not) | LR-, KNN-, SVM-, MLP-, RF-, XGBoost-, LightGBM-based radiomics model | The ExtraTrees model achieved an AUC of 0.787, while the DenseNet121 DL model reached 0.766. The combined Thy-DL-Radiomics model showed improved diagnostic accuracy, with an AUC of 0.839 in internal validation and 0.789 in external validation |
| Ruqian et al. 2024 [48] | US | US data from 528 HT combined PTC patients | Classification (LNM or not) | RF, SVM, LightGBM, KNN, and XGBoost models | All five ML models performed well, with AUCs from 0.798 to 0.921. LightGBM and XGBoost achieved the highest performance, surpassing other models |
| H. Du et al., 2024 [49] | US | US data 817 patients | Classification (Malignant or benign) | Radiomics signature, DL signature, Clinical US semantic signature | The combined model achieved AUCs of 0.947, 0.917, and 0.929, with the comprehensive nomogram showing the best clinical utility. AUCs for differentiating benign and malignant thyroid nodules ranged from 0.714–0.752 for junior physicians to 0.891–0.908 for experts. With DL radiomics assistance, these improved to 0.858–0.923, 0.888–0.944, and 0.912–0.919, respectively |
| H. Zhao et al., 2024 [50] | US | US data from 75 lymph nodes | Classification (CLNM or non-CLNM) | SVM-, DT, and LDA-based radiomics model | The SVM model had the best diagnostic efficacy, with a sensitivity of 93.0%, a specificity of 93.8%, and an accuracy of 93.3%. |
| Abbreviations: US: Ultrasound, CT: Computed tomography, MRI: Magnetic Resonance Imaging, LN: Lymph Node, PTC: Papillary thyroid carcinoma, AUC: Area under the curve, LR: Logistic Regression, RF: Random Forest, KNN: k-nearest neighbor, SVM: Support Vector Machine, DT: Decision Tree, NB: Naive Bayes, SVC: Support Vector Classifier, ETE: extrathyroidal extension, LNM: Lymph Node Metastasis, CLNM: Central Lymph Node Metastasis, FDG-PET: Fluorodeoxyglucose Positron Emission Tomography, CNN: Convolutional Neural Network, TL: Transfer Learning, DLRT: Deep Learning Radiomics Technique, PNN: Probabilistic Neural Network, ANN: Artificial Neural Network, MRMR: Minimum Redundancy Maximum Relevance, USR: Unsupervised Representation Learning, NG: Nodular Goiter, TN: Thyroid Nodule, XGBoost: Extreme Gradient Boosting Machine, GLRLM-RLNU: Gray Level Run Length Matrix - Run-Length Non-Uniformity, GLZLM-GLNU: gray-level zone length matrix - Run-Length Non-Uniformity, HT: Hashimoto's thyroiditis, LDA: linear discriminant analysis, DTC: Differentiated Thyroid Carcinoma, AB: adaptive boosting, LDA: linear discriminant analysis | | | | | |